\documentclass{aa} 
\usepackage[flushleft]{threeparttable}
\usepackage{txfonts}
\usepackage{xcolor}
\usepackage{lscape}
\usepackage{xcolor}
\usepackage{natbib}
\usepackage{enumitem}   
\usepackage{graphicx}
\usepackage{amssymb}
\usepackage{txfonts}
\usepackage[colorlinks=true,allcolors=blue]{hyperref}

\begin{document}

   \title{Physical properties of circumnuclear ionising clusters. \\ IV. NGC\,1097}


   \author{Sandra Zamora \inst{\ref{inst:SNS}}\fnmsep\thanks{E-mail: \href{mailto:sandra.zamoraarenal@sns.it}{sandra.zamoraarenal@sns.it}} 
          \and
          Angeles I. Díaz \inst{\ref{inst:1},\ref{inst:2}}
          \and
          Asier Castrillo \inst{\ref{inst:1},\ref{inst:OAN}}
          }

    \institute{Scuola Normale Superiore, Piazza dei Cavalieri 7, I-56126 Pisa, Italy\label{inst:SNS}
        \and 
        Departamento de Física Teórica, Universidad Autónoma de Madrid, 28049 Madrid, Spain\label{inst:1}
        \and
        CIAFF, Universidad Autónoma de Madrid, 28049 Madrid, Spain\label{inst:2}
        \and
        Observatorio Astronómico Nacional (IGN), C/Alfonso XII, 3, 28014 Madrid, Spain
        \label{inst:OAN}}
        

 
  \abstract
  {
  The circumnuclear star-forming ring of the barred spiral galaxy NGC 1097 provides a unique laboratory to study star formation under extreme conditions. This work aims to derive the physical properties of the circumnuclear star-forming regions (CNSFRs) using MUSE integral field spectroscopy observations. A total of 24 individual ionised HII are identified and analysed within its ring, which spans from $\sim$385 pc to $\sim$1.3 kpc. 
  Despite the complex nuclear activity, all HII regions are found to be purely photoionised.
  Directly derived abundances reveal supersolar metallicities, with the highest one exceeding five times the solar value (12+log(S/H) = 7.875 $\pm$ 0.353, T$_e$([SIII]) = 3912 $\pm$ 567 K), and representing the highest abundance reported to date.
  In this high-metallicity regime, we find a break in the ionisation parameter–[SII]/[SIII] relation, which can be explained by changes in the ionisation structure and line emissivities, as confirmed by photoionisation models that successfully reproduce the observed emission-line ratios.
  Our results also indicate that the local gas supply regulates the star formation activity within the ring, with the young stars ionising 8\% of the total gas in the ring. Furthermore, our findings support a propagating starburst scenario, originating in the galaxy nucleus and extending towards the ends of the bar and into the circumnuclear ring through bar-driven shocks, this being consistent with the results of previous multi-wavelength studies. 
  Finally, we likely detect optical signatures associated with one of the two known jets in this galaxy. This finding, together with the radio core emission previously found at sub-parsec scales, reflects the presence of feedback processes operating even on small galactic disc scales.
  }


   \maketitle
%

\section{Introduction}
\label{sec:introduction}

Galactic nuclei are often thought to be the sites of energetic and violent phenomena related to the presence of central massive black holes which are intimately connected with the structure, origin, and evolution of galaxies, and the study of the physical conditions of the gas in the circumnuclear regions of galaxies can greatly help our understanding of the existing connections between nuclear activity and star formation processes. This gas represents probably the most evolved gaseous system for which we can measure the chemical abundances in a direct manner; these abundances can provide important information about the chemical evolution of galactic nuclei and also serve as tests of theories of nucleosynthesis and stellar evolution. On the other hand, chemical abundances can affect star formation, stellar structure and evolution, and dust particles and these, in turn, can modify the central regions of galaxies.

This is the fourth article of a series  in which the peculiar conditions of star formation in circumnuclear regions are studied using  the full spectral region observed, from 4800 to 9300 \AA , provided by archival data obtained with the MUSE spectrograph attached to one of the ESO VLT. As already explained in the first paper of this series \citep{ngc7742paper}, the usually high abundances of the gas in circumnuclear star forming regions (CNSFR) and their low excitation produce very weak nebular [OIII] lines,  difficult to measure with confidence, thus making practically impossible the use of these lines customary employed for abundance analysis. However, the extended wavelength range to the red covered by MUSE allows the use of auroral and nebular forbidden emission lines of sulphur which has proven to be an excellent abundance tracer for the characterisation of HII regions and ionising clusters  at high metallicities \citep{2022MNRAS.511.4377D}.

NGC 1097 (also known as Arp 77 and Caldwell 67) is a southern hemisphere barred spiral galaxy which displays several remarkable peculiarities.  Discovered back in the late 18th century by William Herschel using  his 18.7 inch reflector in England since, at a latitude of -30 degrees, the source was very bright and visible a few degrees above horizon.  
Its nuclear region was shown by \citet{Burbidge1960} and \citet{Sersic1965} to consist of a bright core surrounded by an almost complete ring of about 1.5 kpc diameter where most of the galaxy’s star formation is currently taking place, as evidenced by the presence of a good number of HII regions. This star formation is thought to be caused by inflows of gas along the galaxy bar. The ring was studied spectroscopically during the last century using a variety of instrumentation in different wavelength ranges (see, for example, \citet{Meaburn1981,Talent1982,Osmer1974} in the optical; \citet{Telesco1981} , in the infrared; and \citet{Wolstencroft1984} at radio frequencies). 

More recently, \citet{Beirao2010} have studied the physical properties of the interstellar medium (ISM) in the ring using spatially resolved far-infrared (IR) observations made with the PACS spectrometer on board the Herschel Space Observatory. Maps of the IR emission lines of oxygen, nitrogen and carbon: [OI] 63 $\mu$m, [OIII] 88 $\mu$m, [NII] 122 $\mu$m, [CII] 158 $\mu$m and [NII] 205 $\mu$m along the ring are presented, showing rapid rotation ($\sim$220 km/s).  

The galaxy shows evidence of interaction with one close satellite, NGC 1097A, distorting its disc and has another satellite, NGC 1097B, discovered by its HI emission, which seems to be a dwarf irregular.  NGC 1097 also shows faint optical jet-like features directed radially from the nucleus and extending out to projected distances as large as 90 kpc, which were discovered by \citet{Wolstencroft1975}. According to \citet{Carter1984} the colours of the jets were consistent with being composed of late-type stars and supposed to be stellar streams resulting from past minor mergers, although some authors suggest that a relationship between the mild activity of the galaxy nucleus and the optical jets might exist \citep{Storchi-Bergmann1993}. 
Additionally, there is a radio core emission and the possible presence of a jet at sub-parsec scale \citep{Mezcua2014,Orienti2010}, although \citet{Prieto2005,Prieto2019} do not see evidence for collimated large-scale radio emission.

\begin{table}
\caption{NGC 1097 global properties.}
\label{tab:galaxy characteristics}
\begin{tabular}{cc}
\hline
Galaxy & \href{https://ned.ipac.caltech.edu/byname?objname=ngc1097&hconst=67.8&omegam=0.308&omegav=0.692&wmap=4&corr_z=1}{NGC1097} \\ \hline
RA J2000 (deg)$^a$ & 41.579375\\ 		
Dec J2000 (deg)$^a$ & -30.274889\\ 
Morphological type & SB(s)b\\  	
Nuclear type & Sy 1 \\
z & 0.00424\\ 
Distance (Mpc)$^b$ & 14.50 \\
Scale (pc/arcsec)$^c$ & 80\\ 	
\hline
\end{tabular}
\begin{tablenotes}
\item $^a$ \citet{2006AJ....131.1163S}.
\item $^b$ \citet{2000ApJ...529..786M}.
\item $^c$ Virgo Infall scale.
\end{tablenotes}
\end{table}
 
In this work, we analyse the circumnuclear environment of  NGC 1097 using publicly available MUSE IFS observations \citep{MUSE} which characteristics are listed in Table \ref{tab:galaxy characteristics}. These observations reveal the prominent circumnuclear star-forming ring, with a diameter $\sim$ 1.5 kpc, showing intense  star formation. As mentioned above, high-resolution optical, infrared, and radio observations reveal the ring as a bright assembly of massive young star clusters, each with masses of $\sim$ 10$^5$ M$_\odot$ at all the ages. 
Using IR to optical high angular resolution observations down to parsec scales in \citep{Prieto2019} they resolved more than 300 young stellar clusters in the ring. The stellar population analysis at cluster level reveal four star cluster generations, occurring at 4 Myr, $\sim$30 Myr, $\sim$ 50-60 My and $\sim$ 90 Myr.

Streamers of gas are observed flowing further inward toward the central black hole. They have been observed in HI and dust and they are channing matter from the outskirts of the galaxy straight to the central parsec, to feed the circumnuclear ring and the BH \citep{Prieto2005,Prieto2019,Ondrechen1989}.

Due to its proximity, brightness, barred structure, active nucleus, and bright circumnuclear star-forming ring, NGC 1097  constitutes an excellent case of study for understanding how bar-driven inflows can build and sustain nuclear rings, as well as their role as starburst reservoirs and as channels feeding the galactic center in the so called  starburst–AGN connection .

In Section 2, we describe the observations; our analysis and the results concerning  the ionised gas are presented in Section 3, including the emission line and continuum maps of the circumnuclear ring, and the selection, emission line measurements and chemical abundances of the HII regions associated with it. In Section 4, we discuss the results in the context of circumnuclear regions in other galaxies and using a multi-wavelength approach. Finally, Section 5 summarizes this work and presents our conclusions.

\section{Observations}
\label{observations}

NGC 1097 was observed on 2017 June 14 as a part of the ESO Programme 097.B-0640(A). The observations consisted of eight exposures summing up to a total exposure time of 3840\,s, with 1\,arcsec offsets in declination and different rotation angles between exposures. Sky frames were acquired following the target observations to allow for proper sky subtraction. The median seeing was 0.525\,arcsec. Data reduction was carried out by the ESO Quality Control Group using an automated process with MUSE pipeline version 1.6.4 \citep{MUSEpipeline}, following the procedures explained in \citet{ngc7742paper}.

\section{Data analysis and ionised gas results}
\label{sec:results}
We analysed the data following the methodology described in \cite{ngc7742paper} \citep[see also][]{NGC7469paper}. However, due to the particular complexity of this galaxy, our study focuses exclusively on the ionised gas and does not address the stellar cluster population properties. The main steps of the analysis are: (i) performing 2D maps for different emission lines and continuum bands; (ii) selecting HII regions from the H$\alpha$ emission line map; (iii) extracting each region spectrum and measuring the available emission lines; (iv) deriving chemical abundances for each of the CNSFRs. In what follows, only specific details introduced for this galaxy are explained.

\subsection{Emission line and continuum maps}
\label{emmision maps}

\begin{figure*}
\includegraphics[width=\textwidth]{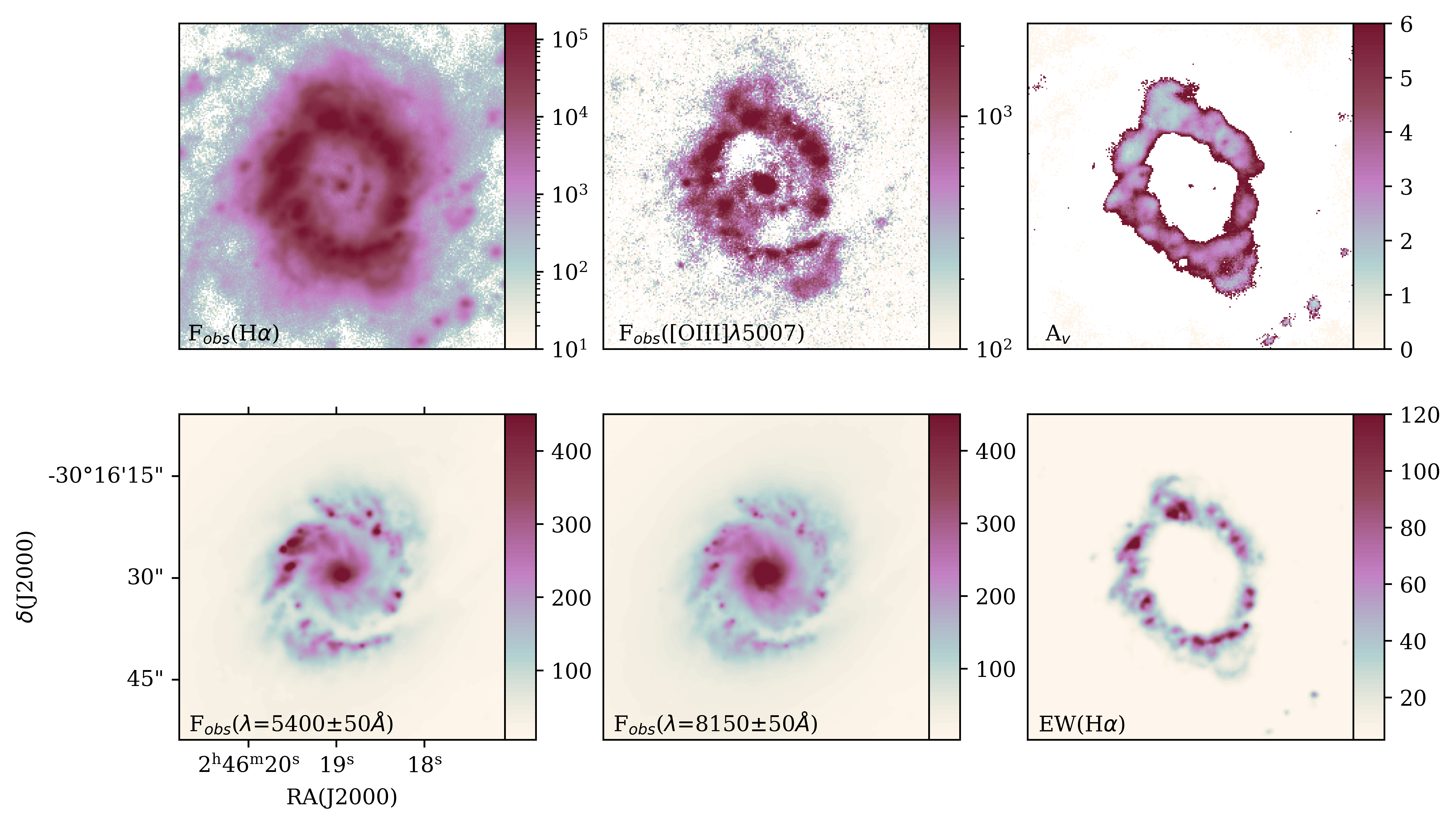}
\caption{From left to right and top to bottom: maps of the observed H$\alpha$ and [OIII]$\lambda$5007 \r{A} emission line fluxes (in units of 10$^{-20}$ erg/s/cm$^2$ and logarithmic scale); A$_V$ extinction (in magnitudes); observed continuum fluxes in the blue and red parts of the spectrum (5400 \AA\ and 8150\AA\ respectively), in units of 10$^{-17}$ erg/s/cm$^2$ and logarithmic scale); and EW($H\alpha$) in \r{A}. Orientation is North up and East to the left.}
\label{fig:Ha_OIII_map}
\end{figure*}

From the data cube, we constructed two-dimensional maps for various emission lines and two continuum bands. The top-left panel of Fig.~\ref{fig:Ha_OIII_map} shows the spatial distribution of the observed H$\alpha$ flux, revealing ionised HII regions surrounded by diffuse gas remaining into the ring, which forms arcs and stream-like structures. The nucleus and the inner ring of the galaxy are clearly distinguished, and several smaller ionised regions are also visible, located within the ring near the central part of the galaxy.

The top-central panel of the same figure presents the [OIII]$\lambda$5007 \AA\ emission map. This emission is dominated by the nucleus, and the ionised regions, with the ring having a less diffuse morphology compared to the H$\alpha$ emission. Additionally, an arc in the lower right part of the ring is clearly visible, which corresponds with the location of one of the jets observed in this galaxy \citep{Wolstencroft1975,Arp1976,Lorre1978,Phillips1984}.

The top-right panel shows the extinction map derived from the combination of the H$\alpha$ and H$\beta$ maps, using the Galactic extinction law of \citet{reddening}, a specific attenuation of $R_{V} = 2.97$, and the theoretical H$\alpha$/H$\beta$ ratio of 2.87 from \citet{Osterbrock2006} ($n_{e} = 100$ cm$^{-3}$, $T_{e} = 10^{4}$ K, case B recombination). In this map, the inner ring is clearly visible, showing lower extinction in its upper part ($\sim$1.5 mag) compared to its lower one ($\sim$3 mag), likely due to the inclination of the ring.

The two bottom-left panels show the maps of the observed continuum flux at blue (5400 \AA) and red (8150 \AA) wavelengths. In both maps, the continuum emission originates from the center of the galaxy, and a two armed structure is clearly identified, along with a bar that becomes more prominent in the redder continuum likely because it is formed by an old stellar population. The dominant continuum emission comes from the young star ionising clusters, which show more emission at blue wavelengths.  

Finally, the bottom-right panel of Fig. \ref{fig:Ha_OIII_map} shows the map of the H$\alpha$ equivalent width (EW) in \AA. Notably, two distinct streams are evident: one coinciding with the position of the [OIII] arc and another on the opposite side. Its origin is likely associated with an interaction between the interstellar medium (ISM) and the jets, which are located exactly in these directions, and can compress or shock the surrounding gas, enhancing its ionisation and producing the observed emission features. All circumnuclear regions show EW(H$\alpha$) values greater than 100~\AA, consistent with the presence of very massive stars formed in a recent star-formation episode within the last 10~Myr.

\subsection{HII region selection}
\label{sec:segmentation}

The HII region selection method is described in detail in \citet{ngc7742paper}. First, we identified the projected size of the ring from the H$\alpha$ pixel by pixel radial profile, which extends from 4.8 to 16.4 arcsec \citep[$\sim$ 385 pc - 1.3 Kpc assuming a distance of 16.6 Mpc,][]{2000ApJ...529..786M}. 

Then, we used an iterative procedure applied to the H$\alpha$ emission line maps, which detects high intensity clumps and then adds adjacent pixels according to several input parameters: the level of diffuse gas emission, the relative flux intensity of each region with respect to its central peak, and the maximum and minimum extents of the regions, determined according to their typical projected sizes and the point spread function (PSF) of the observations respectively. 

Finally, we imposed two quality control requirements to the integrated spectra extracted from each selected region to ensure their physical meaning and the star formation origin of the emission: (i) EW(H$\alpha$) > 6 \AA\ \citep{Sanchez2015}, and (ii) 2.7 < H$\alpha$/H$\beta$  < 6.0 \citep[][n$_e$ = 100 cm$^{-3}$, T$_e$ = 10$^4$ K]{Osterbrock2006}.

\begin{figure}
\centering
 \includegraphics[width=0.9\linewidth]{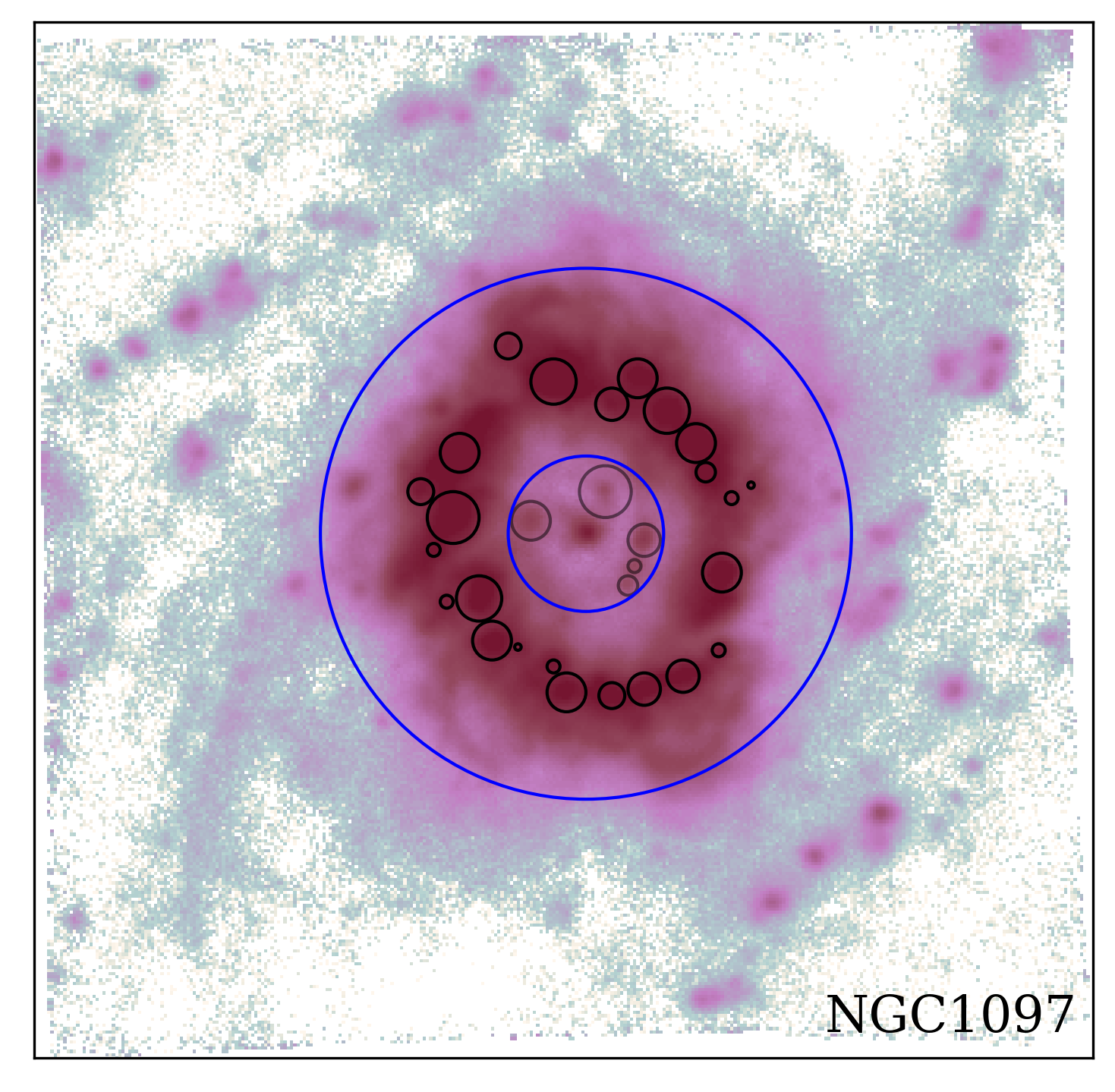}
 \caption{HII regions selected using our segregation program on the H$\alpha$ observed emission line  map. The aperture used to define the ring for the segregation of the regions is overplotted in blue. Logarithmic color scale. Orientation is north up, east to the left. The physical scale is represented at the bottom left corner of the map.}
 \label{fig:ring_ha_profile}
\end{figure}

At the end of the procedure, we identified a total of 24 HII regions in the ring. We repeated the procedure in the inner circumnuclear region outside the ring and we identified 5 additional regions. Figure \ref{fig:ring_ha_profile} shows the HII regions selected and Table \ref{tab:seleccion} summarizes their properties, including the position of each region relative to the galaxy center, its size, and the observed integrated H$\alpha$ flux.

We identified our HII regions with those selected by \citet[][K00]{Kotilainen2000} using the Br$\gamma$ emission in the infrared. Their observations were performed with UKIRT–IRCAM3 photometry in the JHK bands, with a seeing of FWHM = 0.6–0.7 arcsec and a pixel scale of 0.281 arcsec/pix. Ten of our selected regions show corresponding IR emission, with one of them being a double identification in their study. We repeated the identification for the emission regions reported by \citet[][H97]{Hummel1987}, who used radio continuum emission at 1.465 GHz from Very Large Array observations with a spatial resolution of 2.5 arcsec. Seven of our HII regions have emission in radio continuum. The correspondence between our identified regions and those reported by the authors is presented in Table~\ref{tab:multiw}.

\subsection{Emission line measurements and uncertainties}
\label{sec:line measurements}

We extracted the spectrum of each region by integrating the flux within its corresponding aperture and we measured the intensity of their emission lines following the procedure described in \citet{ngc7742paper}.We requested a S/N ratio larger than 3 for the strong emission lines: H$\beta$ and H$\alpha$ Balmer lines; [OIII]$\lambda\lambda$ 4959,5007 \AA , [NII]$\lambda\lambda$ 6548,84 \AA , [SII]$\lambda\lambda$ 6716,31 \AA , and [SIII]$\lambda$ 9069 \AA\ forbidden lines. In addition, we measured the weak [SIII]$\lambda$ 6312 \AA\ auroral line with S/N > 1. We did not detect the [SIII]$\lambda$ 9069 \AA\  forbidden line in the  inner regions outside the ring, whereas the auroral [SIII] line was detected in 11 ring regions. 

We corrected the measured line intensities for dust extinction using the coefficient c(H$\beta$), derived from the observed H$\alpha$ and H$\beta$ Balmer ratio. A simple screen dust distribution was assumed, with identical extinction applied to both the emission lines and the stellar continuum. We adopted the Galactic extinction law of \citet{reddening}, with a specific attenuation of R$_{V}$ = 2.97. A theoretical H$\alpha $/H$\beta $ ratio of 2.87 was used \citep[n$_e$ = 100 cm$^{-3}$ and T$_e$ = 10$^4$ K,][]{Osterbrock2006}.

\begin{figure*}
\centering
\includegraphics[width=\linewidth]{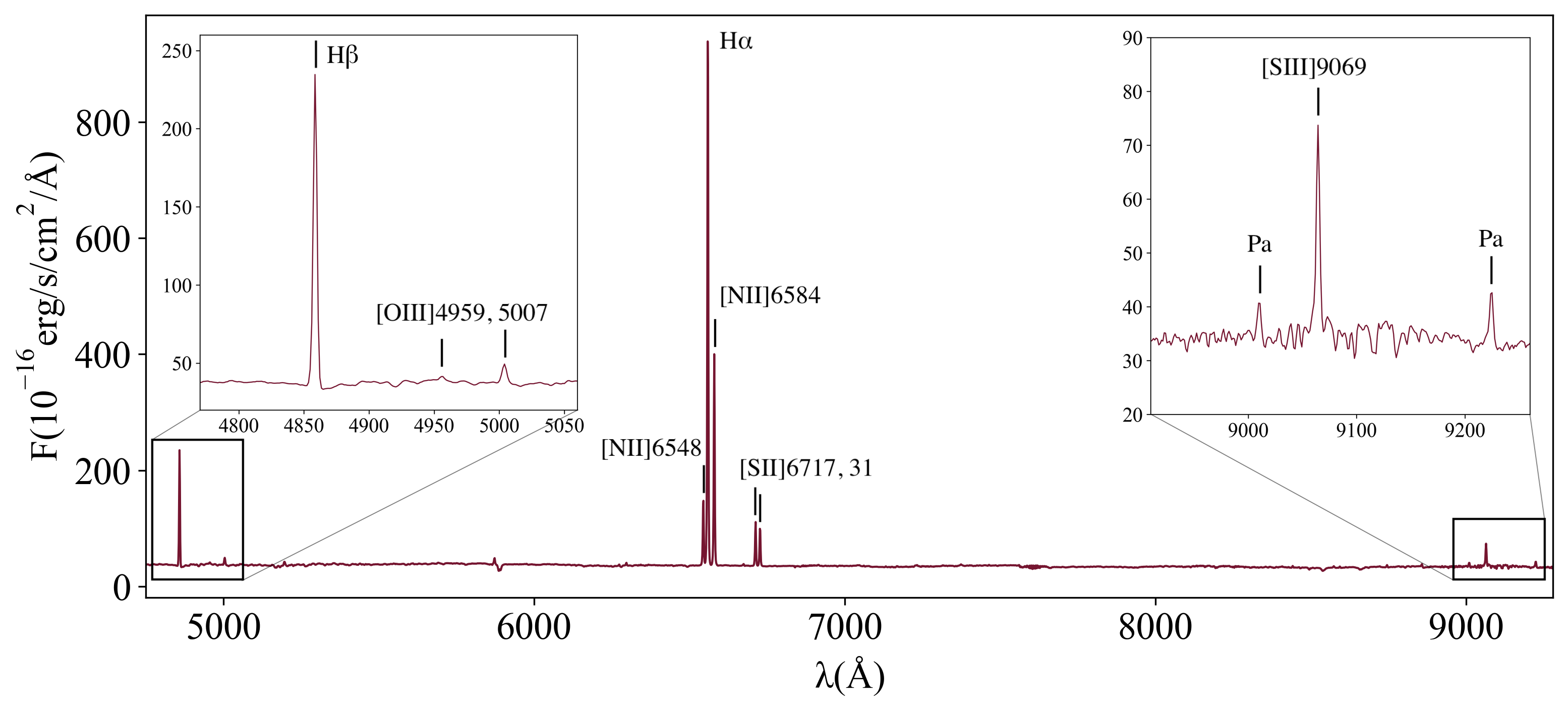}
\caption{Extracted spectrum of region R2.} 
\label{fig:esp}
\end{figure*}

A typical HII region spectrum is shown in Fig. \ref{fig:esp}, illustrating that in high metallicity environments such as the central regions of galaxies, the [OIII]$\lambda\lambda$ 4959,5007 \AA\ emission lines, commonly used to measure chemical abundances, are almost undetectable, while the nebular [SIII]$\lambda$ 9069 \AA\ line is clearly detected. Table \ref{tab:lines} presents the reddening-corrected emission line intensities of the strong lines relative to H$\beta$,  together with the corresponding reddening constant for all the ring regions analysed.

\subsection{Chemical abundances}
\label{sec:abundances}

We measured the CNSFR metallicities using their sulphur abundances following the methodology described in \citet{2022MNRAS.511.4377D}, which is particularly well suited for the MUSE wavelength range, using the sulphur lines as a tracer of abundances. Its main advantages are: (i) reddening effects are reduced due to the longer wavelengths involved, (ii) sulphur does not appear to be depleted in diffuse clouds \citep{2021A&A...648A.120R}, and (iii) the [SIII]$\lambda 6312$ \AA\ auroral line can be detected and measured up to at least solar abundances \citep{Diaz2007}, as those expected in the central regions of galaxies.

We measured this line with a S/N ratio higher than 1 in approximately 46\% of the HII regions, consistent with the detection fractions reported in previous studies \citep[35--45\%,][]{ngc7742paper,NGC7469paper}. For these regions, total sulphur abundances have been derived using the direct method described in \citet{ngc7742paper}. In this approach, the [SIII] electron temperature, T$_e$([SIII]), is calculated from the ratio of the nebular to auroral lines of this ion. We obtained a median value of T$_e$([SIII]) $\sim$ 0.5486 $\times$ 10$^4$ K. 

The equation used to determine this temperature exhibits only a very weak dependence on the electron density. Nevertheless, we employed PyNeb \citep{pyneb} to compute the electron densities of the CNSFRs from the [SII]$\lambda \lambda$ 6717,6731 \AA\ doublet ratio. The derived electron densities within the ring are low and confined to a narrow range centred around n$_e$ $\sim$ 240 cm$^{-3}$, with a median value of 236 cm$^{-3}$ and a standard deviation of 67 cm$^{-3}$, corresponding to the lower limit of densities measurable with these lines. Therefore, the weak dependence of the temperature calculation on the density is not relevant in our case, since T$_e$([SIII]) increases by only 3\% for n$_e$ values between 100 and 1000 cm$^{-3}$ \citep{Perez-Montero2017}.

Assuming a single-zone approximation where T$_e$(S$^+$) $\sim$ T$_e$(S$^{++}$) = T$_e$([SIII]), representing the characteristic electron temperature of the whole nebula, we derived the total sulphur abundances for these CNSFRs. We obtained a median close to the solar value 12+log(S/H) = 7.11. The highest measured abundance reaches (12+log(S/H) = 7.88 $\pm$ 0.35), corresponding to more than five times the solar value. The two ionic species present, S$^{+}$ and S$^{++}$, contribute approximately 60 per cent each to the total abundance (the S$^{+}$/S ionic fraction take values between $\sim$ 45-70 \%). Table \ref{tab:sulfur_measurements} lists these results. 

For the remaining regions, we derived empirical sulphur abundances using the S$_{23}$ parameter and the calibration of \citet{2022MNRAS.511.4377D}. The median sulphur abundance derived using this empirical method is 12+log(S/H) = 6.42, which is significantly lower than the values obtained from the direct method. However, we note that the empirical S$_{23}$ calibration may be affected by the effective temperature of the ionising radiation, particularly at high abundances. As discussed by \citet[see Sec.~5.3]{2022MNRAS.511.4377D}, lower effective temperatures can shift the calibration towards higher abundances by up to 14\% in the logarithm. Consequently, given the high sulphur abundances found in the CNSFRs of this ring from the direct method determination, it is likely that we have underestimated the abundances in the regions where the auroral line was not detected. The resulting sulphur abundances for all objects in our sample are listed in Table \ref{tab3}.  

\section{Discussion}
\label{discussion}

\subsection{Ionisation nature}
\label{sec:nature}

According to \citet[][see Fig. 11]{Kolcu2023}, the emission line ratios in the ring are consistent with the predictions of star forming models. However, this galaxy hosts a  low-luminosity active galactic nucleus (LINER) with a bolometric luminosity of L$_{bol}$ = 6.2 $\times$ 10$^{41}$ erg/s \citep{prieto2010,Fernandez-ontiveros2023}, L$_{bol}$ = 8.6 $\times$ 10$^{41}$ erg/s \citep{Nemmen2006}, as well as four optical jets extending radially from the nucleus \citep{Wolstencroft1975,Arp1976,Lorre1978}. These jets occur in two opposite pairs and reach projected distances of up to $\sim$ 90 kpc. In our optical spectroscopic observations, we detected signatures of two of these jets, suggesting that AGN-driven processes may also contribute to the observed gas excitation in certain regions of the galaxy close to our nuclear ring. Therefore, before analysing the CNSFRs, we first verified if the origin of the emission is purely due to star formation processes.

\begin{figure*}
\centering
\includegraphics[width=0.37\textwidth]{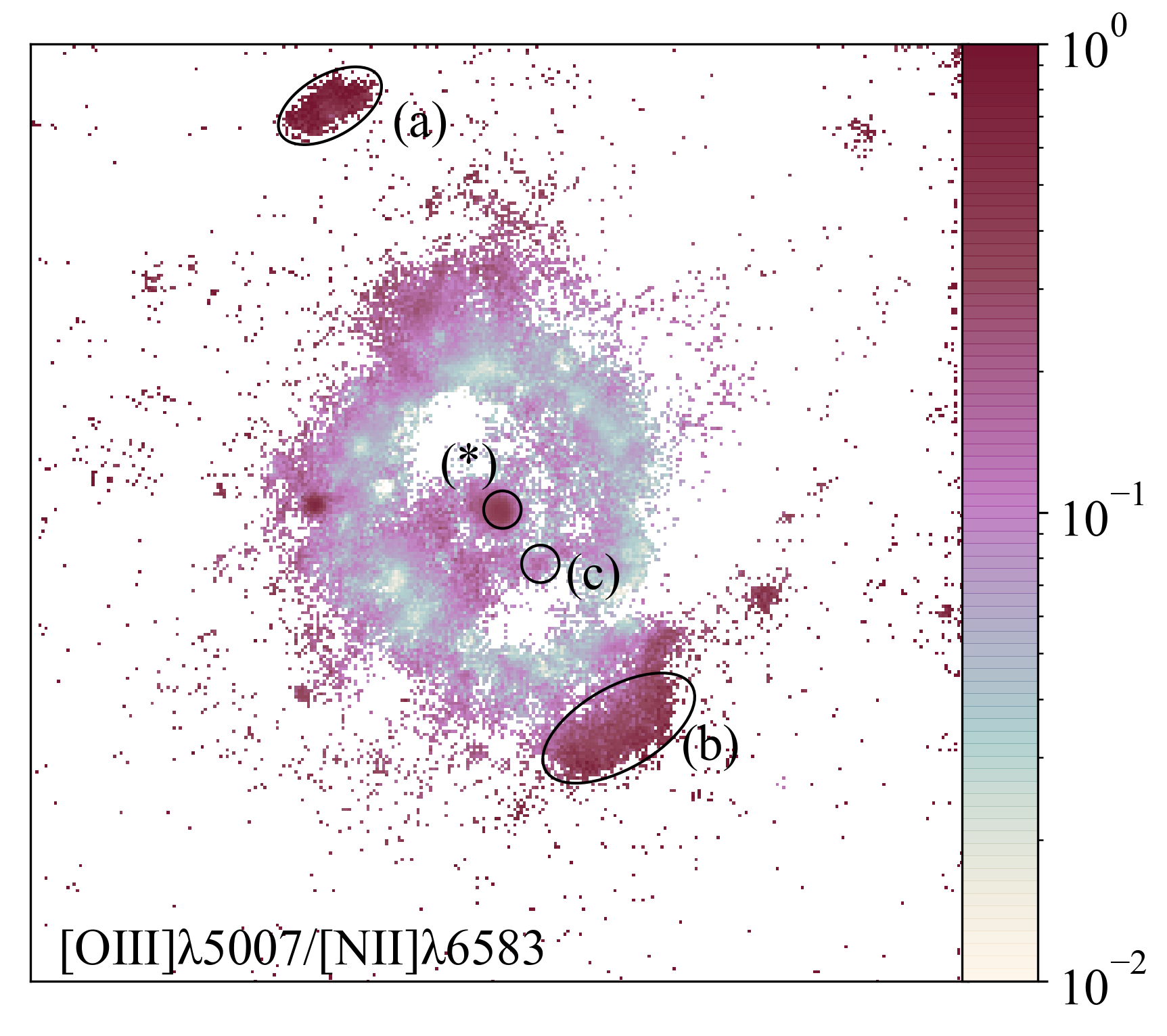}
\includegraphics[width=0.61\textwidth]{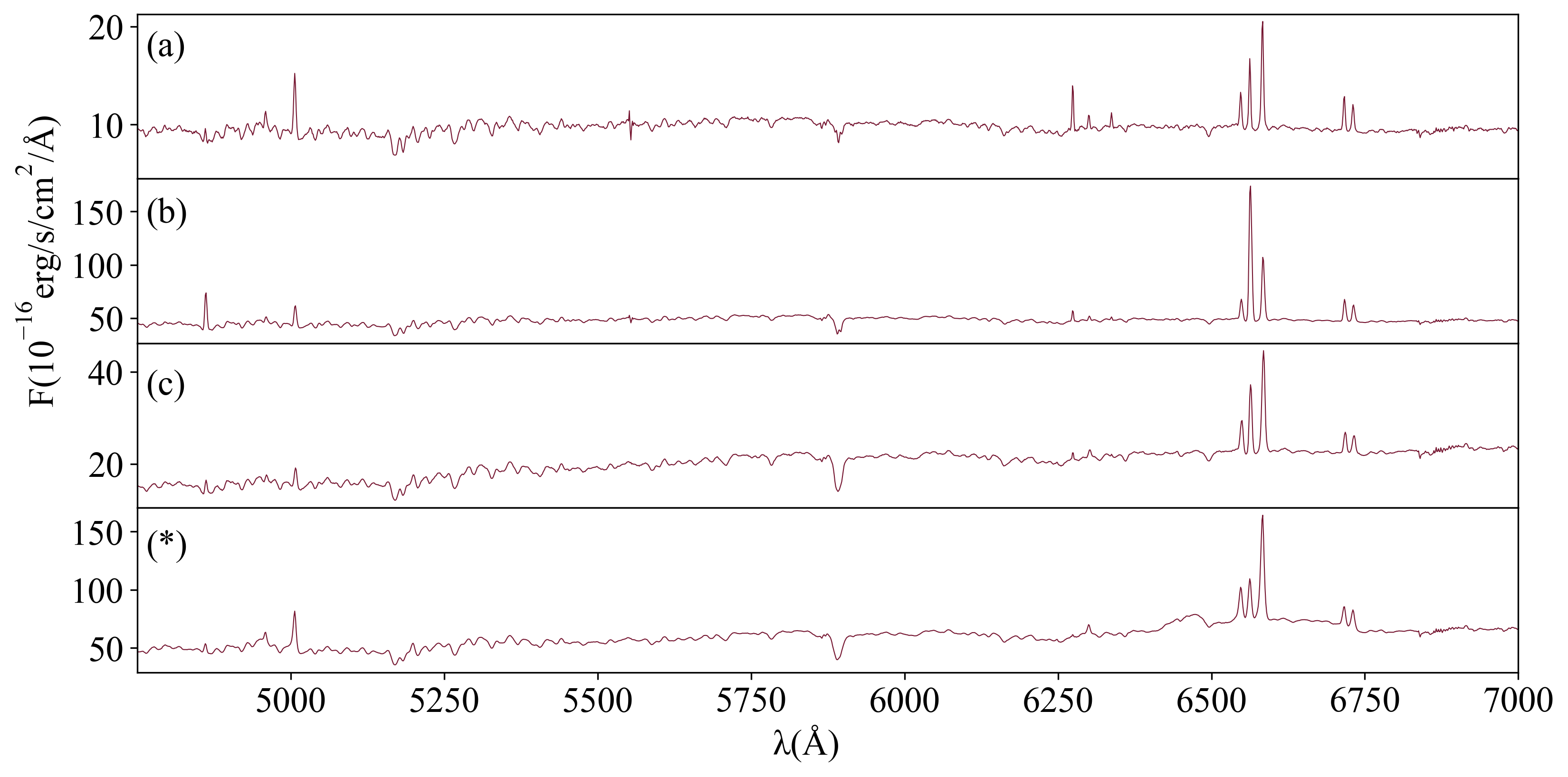}
\includegraphics[width=.9\columnwidth]{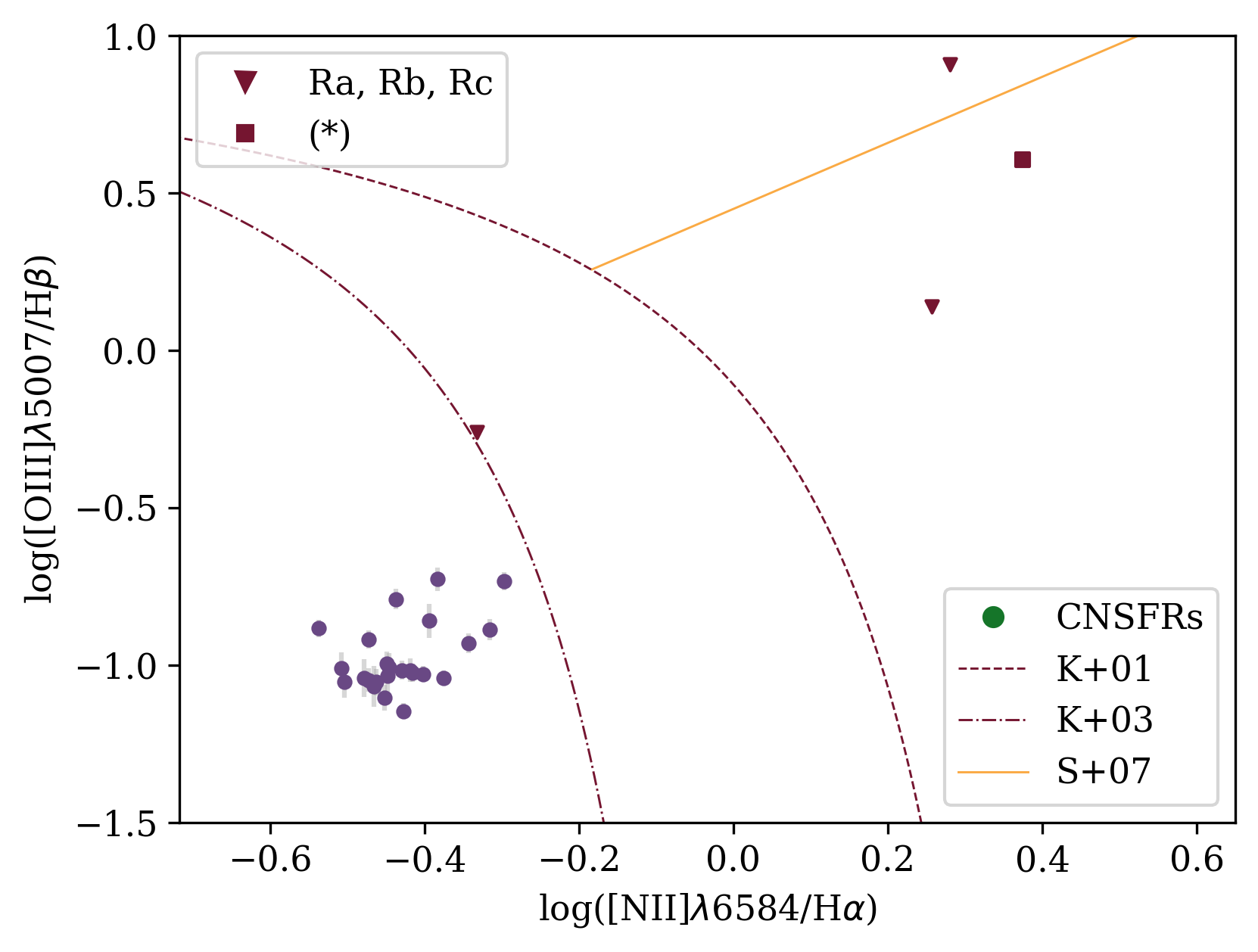}
\includegraphics[width=.9\columnwidth]{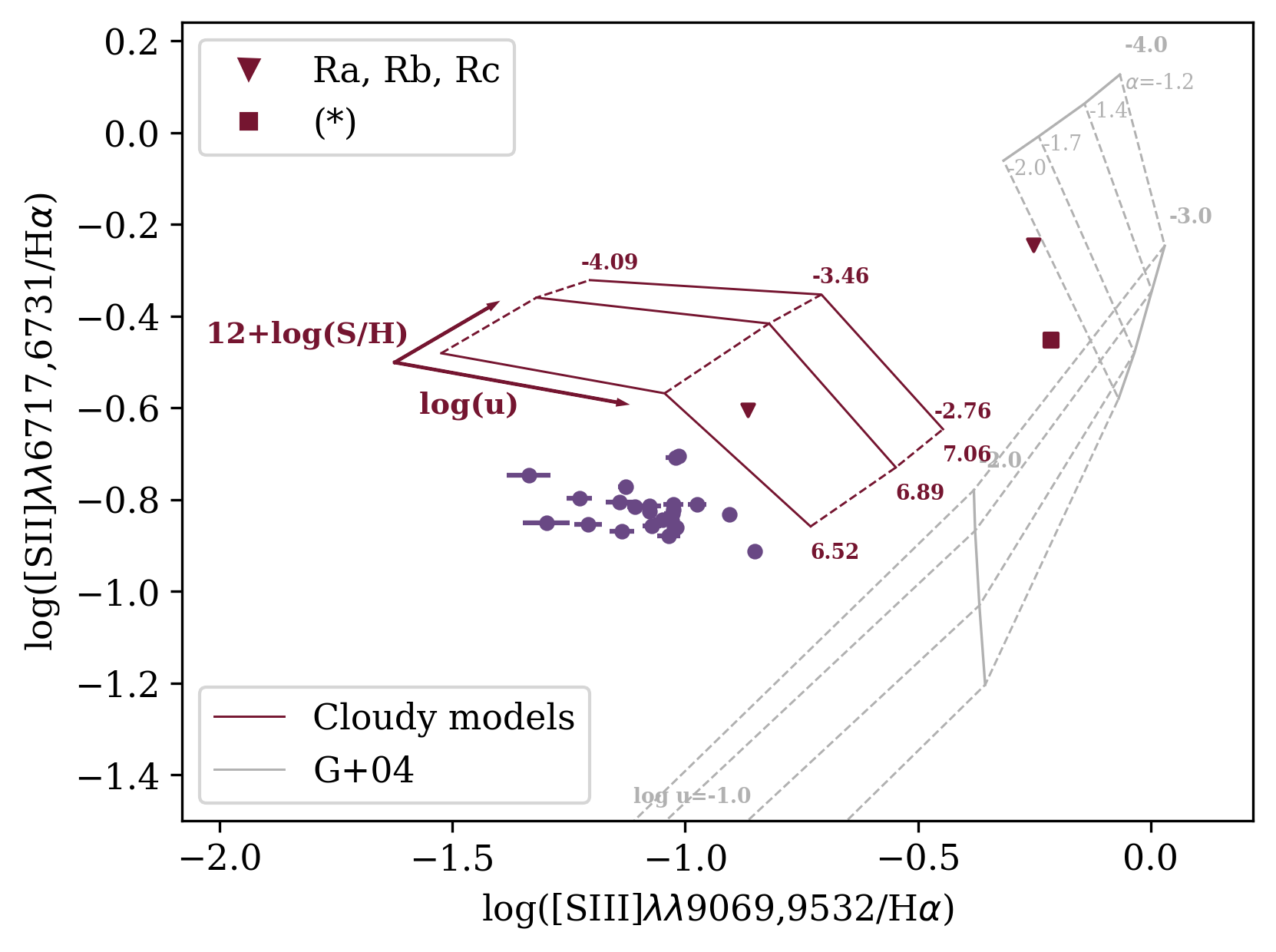}
\caption{Upper left panel: map of the observed [OIII]$\lambda$ 5007 \AA\ /[NII]$\lambda$ 6584 \AA\ ratio. Upper right panel: Emission line spectra of regions contaminated by the AGN emission (a, b, c) and the LINER nucleus of the galaxy (*). Lower left panel: the [OIII]/H$\beta$ vs [NII]/H$\alpha$ diagnostic diagram. Over-plotted, derived separations between LINER/Seyfert \citep[S+07,][]{2007MNRAS.382.1415S} and HII regions \citep[K+01 and K+03,][]{2001ApJ...556..121K, Kauffmann2003} Lower right panel: The [SII]/H$\alpha$ - [SIII]/H$\alpha$ diagnostic diagram. Over-plotted, dust-free AGN photoionisation models \citep[G+04,][]{2004ApJS..153....9G} and star forming models by \citet{ngc7742paper}. }
\label{fig:bpt}
\end{figure*}

The top left panel of Fig.~\ref{fig:bpt} shows the spatial distribution of the  [OIII]$\lambda$ 5007 \AA\ /[NII]$\lambda$ 6584 \AA\ ratio.  Two arc structures with enhanced ratios, consistent with shocked gas, 
are clearly identified in the direction of one of the jet pairs (previously noted in Sec. \ref{emmision maps}). To study these features, we defined two elliptical apertures encompassing the arcs: (a) centred at RA = 116.92 deg, DEC = 293.17 deg, with PA =210 deg, a = 3.64 arcsec, b = 1.96 arcsec; and (b) centred at RA = 209.32 deg, Dec = 93.97 deg, with a = 5.4 arcsec, b = 2.6 arcsec. We also identified a region of high ratio values in the lower clumps withing the circumnuclear ring, defining an additional circular aperture (c) centred at RA = 184.24 deg, Dec = 146.53 deg with radius  r=1.2 arcsec, together with a similar circular aperture at the galaxy centre. The integrated spectra extracted from all these apertures are shown in the top right panels of Fig. \ref{fig:bpt}, where we applied the same fitting procedure described in Sec. \ref{sec:line measurements} to measure the emission lines present in them. The spectrum corresponding to the galaxy nucleus is shown at the bottom panel,  marked with an asterisk, and shows characteristics of a mild activity.

Actually, already more than 50 years ago, \citet{Smith1972} noted that the [NII]/H$\alpha$ intensity ratio changed from less than 1 in the HII regions in the ring to a value of greater than 1 in the nucleus, something that was confirmed by \citet{Meaburn1981} together with the finding of very broad ($\sim$500 km/s) profiles of the optical emission lines that pointed to the galaxy having a Seyfert type 2 nucleus, although some years later \citet{Keel1983} classified it as a LINER. 

Ten years later, \citet{Storchi-Bergmann1993} reported the appearance of a very broad component in the H$\alpha$ and H$\beta$ emission lines, first observed in H$\alpha$ in 1991 November 2, and confirmed 11 months later, with FWZI about 21000 km/s  and a double-peaked profile. The data also suggested the presence of a blue, featureless continuum in the nucleus. The H$\alpha$/H$\beta$ flux ratio indicated that the broad- line region (BLR) was not significantly reddened. This finding pointed out to the presence of a Seyfert 1 nucleus that had not been seen before. The authors discussed different alternatives that could give rise to the observed profile (biconical outflow and an accretion disc) and speculated on the possible relationship between the broad line region and the previously known optical jets. 

\citet{Storchi-Bergmann1995} presented a series of spectroscopic observations to follow up the evolution of the nuclear BLR spanning the period 1991-1994, finding that the broad component of H$\alpha$ had varied significantly, both in flux, decreasing by a factor of two, and in its profile shape, that became more symmetrical than originally observed. While the first effect could be interpreted as consequences of either increased obscuration along the line of sight, or a decline in the ionising continuum, the second  one was difficult to account for and no preferred scenario was proposed. 

At present, the nucleus of NGC 1097 is classified as Seyfert 1 (see top right pannel of Fig. \ref{fig:bpt}). The mass of the supermassive massive black hole in its centre has been estimated by \citet{Onishi2015} from observations of dense molecular gas dynamics traced with HCN(J = 1 - 0) and HC{{O}$^+$}(J = 1 - 0) emission lines yielding a value of 1.4 $\times$ 10$^8$ M$_\odot$. 

Given the variability nature and the complex evolution of the nuclear region, we used the classical BPT diagnostic diagram to test the nature of the CNSFRs as shown in the lower left panels of Fig. \ref{fig:bpt}. The BPT diagnostic diagram \citep{bpt} uses the emission line ratios [NII]$\lambda$ 6584 / H$\alpha$ and [OIII]$\lambda$ 5007/H$\beta$. In this diagram, we also plotted the ratios measured in the spectra of the apertures defined above. The star forming regions occupy the high-metallicity end of the empirical star-forming sequence defined by \citet{Kauffmann2003}, as expected for circumnuclear enriched environments \citep[see also][]{Kolcu2023}. In contrast, the arc structures identified in the [OIII]/[NII] maps clearly deviate from the purely photoionisation area. Instead, they present a shock-driven ionisation contribution placed them into the intermediate zone between the LINER and Seyfert classifications \citep{2007MNRAS.382.1415S}. This result is consistent with the known presence of an active nucleus and extended jets in this galaxy, which may interact with the surrounding interstellar medium and locally enhance the excitation of the gas.

Finally, the lower right panel of the same figure presents an additional diagnostic diagram based on the near-infrared sulphur emission lines, which provide a powerful tool for distinguishing between shock and photoionisation mechanisms \citep{Diaz1985}. This diagnostic is particularly suitable for our regions because it is not sensitive to the N/O ratio, difficult to estimate for CNSFRs, and is almost insensitive to reddening effects. The position of our regions in this plot is compatible with star formation excitation but they exhibit [SII]$\lambda\lambda$6717,31/H$\alpha$ ratios lower than those typically found in other CNSFRs \citep[see][]{ngc7742paper,NGC7469paper}. Interestingly, despite their higher abundances, these regions occupy the lower part of the diagram. This may reflect the double valued behavior of some abundance parameters (e.g., O${23}$, S${23}$), suggesting that our regions lie on the upper branch of these calibrations. 

\subsection{Supersolar CNSFR abundances}\label{sec:supersolar}

The high sulphur abundances derived for the CNSFRs in this study are not isolated cases in the literature. The highest sulphur abundance reported so far was measured in region 11 of NGC 5236, placed in the central region of M83 \citep{Bresolin2005}. For this HII region, the authors derived an electron temperature of T$_e$([SIII]) =  4800 $\pm$ 200 K. Also \citet{2022MNRAS.511.4377D} obtained for this region a temperature of 4858 $\pm$ 338 K, fully consistent with the former value within the quoted uncertainties. The corresponding sulphur abundances reported in these two works are 12+log(S/H) = 7.71 and 7.82, respectively. This remarkable concordance between independent determinations strongly supports the reliability of the sulphur abundance estimates and indicates that such elevated values are realistic in the central, metal-rich environments of galaxies.

In this work, we found one region, R6 with a the lowest measured S$^{++}$ temperature T$_e$([SIII]) = 3912 $\pm$ 567 K. Hence, this region has the highest sulphur abundance known up to date, with a value of 7.875 $\pm$ 0.353 which corresponds to more than five times the solar value. 

We used the code \textsc{CLOUDY} \citep{cloudy} to run photoionisation models in order to further support the reliability of our high sulphur abundances determinations and to deeper understand the underlying physical conditions of these regions. These models were constructed using the physical properties derived for those regions where the auroral [SIII]$\lambda$ 6312 \AA\ line was detected, allowing us to directly compare the predicted emission line flux ratios with our observations. We assumed an ionisation-bounded nebula with a plane-parallel geometry. The computed models adopt ionisation parameter values of log(u) = –3.0 and -2.5, an electron density of $n_e$ = 100 cm$^{-3}$, and metallicities of 12+log(S/H) = 6.5, 7.1, 7.3, 7.4, 7.5, and 7.6, assuming that the stellar metallicity is the same as that of the gas. The nebula is ionised by a young stellar cluster synthesized using the \textsc{PopStar} code \citep{Popstar}, adopting a Salpeter initial mass function \citep{Salpeter1955} with lower and upper mass limits of 0.85 and 120 M$_\odot$, respectively, and including the nebular continuum in a self-consistent way. We selected stellar cluster ages of 4, 5, and 6 Myr in order to cover a range of young stellar populations both with and without significant Wolf–Rayet (WR) emission, which change the effective temperature of the cluster. 

\begin{figure}
\includegraphics[width=\columnwidth]{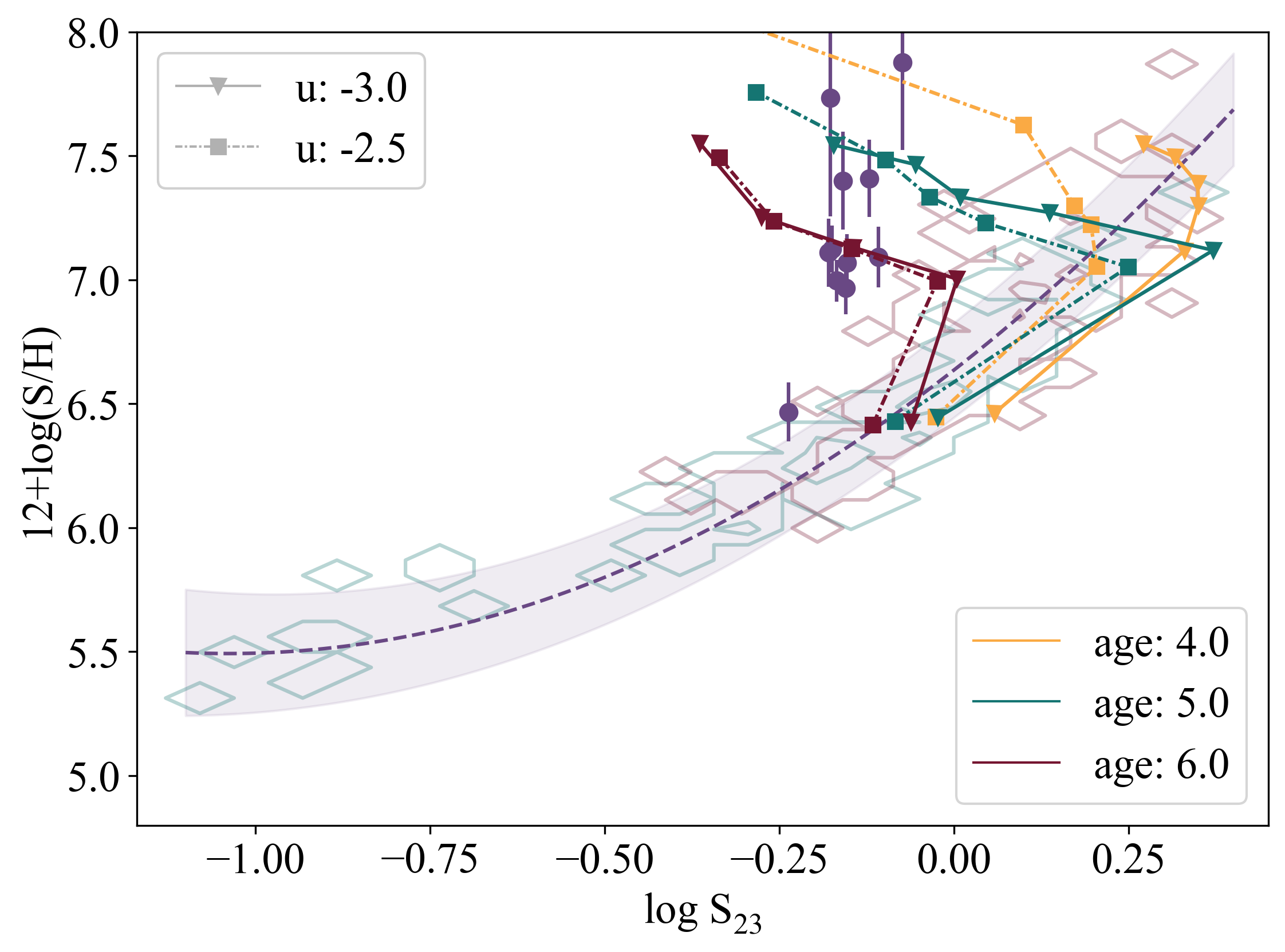}
\caption{The S$_{23}$ abundance calibration from \citet{2022MNRAS.511.4377D}. Red contours correspond to disc HII regions while blue contours correspond to HII galaxies. Purple dots represent the CNFSRs analysed in this work. The Cloudy models described in the text appear superimposed.} 
\label{fig:calibracion}
\end{figure}

Figure \ref{fig:calibracion} shows the empirical S$_{23}$ calibration together with red and blue contours corresponding to data for disc HII regions and HII galaxies respectively from \citet{2022MNRAS.511.4377D}. The individual purple symbols show the measurements of the 11 CNSFRs in this work with auroral [SIII] line detection. The Cloudy models computed for these regions appear superimpose for different ages and ionisation parameters with different colors and line styles. The models reproduce well the observed measurements, with the highest sulphur abundances being consistent with models that assume younger stellar populations. Furthermore, the models show that, at such high metallicities, the behavior of the empirical S$_{23}$ calibration changes in a similar way to the well known turnover of the oxygen R$_{23}$ calibration at subsolar metallicities \citep{Pagel1979}. In this high-abundance regime, the calibration becomes primarily dependent on the effective temperature of the ionising radiation rather than on the ionisation parameter \citep[see Section~5.3 of][]{2022MNRAS.511.4377D}, a trend now supported not only by observations but also by photoionisation models.

Although the derived abundance values may appear extreme, it is important to note that these CNSFRs have been observed previously in the literature. \citet{Phillips1984} obtained long-slit spectra with the Image Photon Counting System (IPCS) at the 3.9 m Anglo-Australian Telescope (AAT), targeting both the circumnuclear ring and the central region of this galaxy (PA = 57$^\circ$, perpendicular to the bar). In these data, the continuum spectrum in the central part of the galaxy is indistinguishable from that of an elliptical galaxy, showing spectral features typical of an evolved, high abundance stellar population (i.e. NGC7145, CaII, band G, MgIb and NaID). The continuum within the ring was found to be very blue, with the Balmer series clearly detected in absorption, in agreement with our CNSFR spectra and consistent with a young population of hot stars.

From the analysis of relative emission-line intensities, \citet{Phillips1984} concluded that the circumnuclear ring regions exhibit properties typical of those in the most metal rich HII regions found in the discs of spiral galaxies,  similar to the regions I and III observed in M83 \citep{Dufour1980}. Using the calibration of \citet{Pagel1981}, they derived oxygen abundances approximately three times the solar value, with log(O/H) = 9.40 and 9.28 for the NE and SW regions respectively (assuming log(O/H)$_\odot$ = 8.92). \citet{Phillips1984} also found a slightly supersolar nitrogen to oxygen ratio, with log(N/O) = 0.77 and 0.87 for the NE and SW regions, respectively. These measurements are consistent with ours in the sense that this work also finds a very high metallicity content in the ionised gas. Moreover, it constitutes an independent determination, as it was based on the ratio ([OII]$\lambda \lambda$3727,3729 + [OIII]$\lambda \lambda$4959,5007)/H$\beta$, which involves bluer wavelengths, lines with different ionisation potentials, and is more strongly affected by dust extinction effects.

\begin{figure}
\includegraphics[width=\columnwidth]{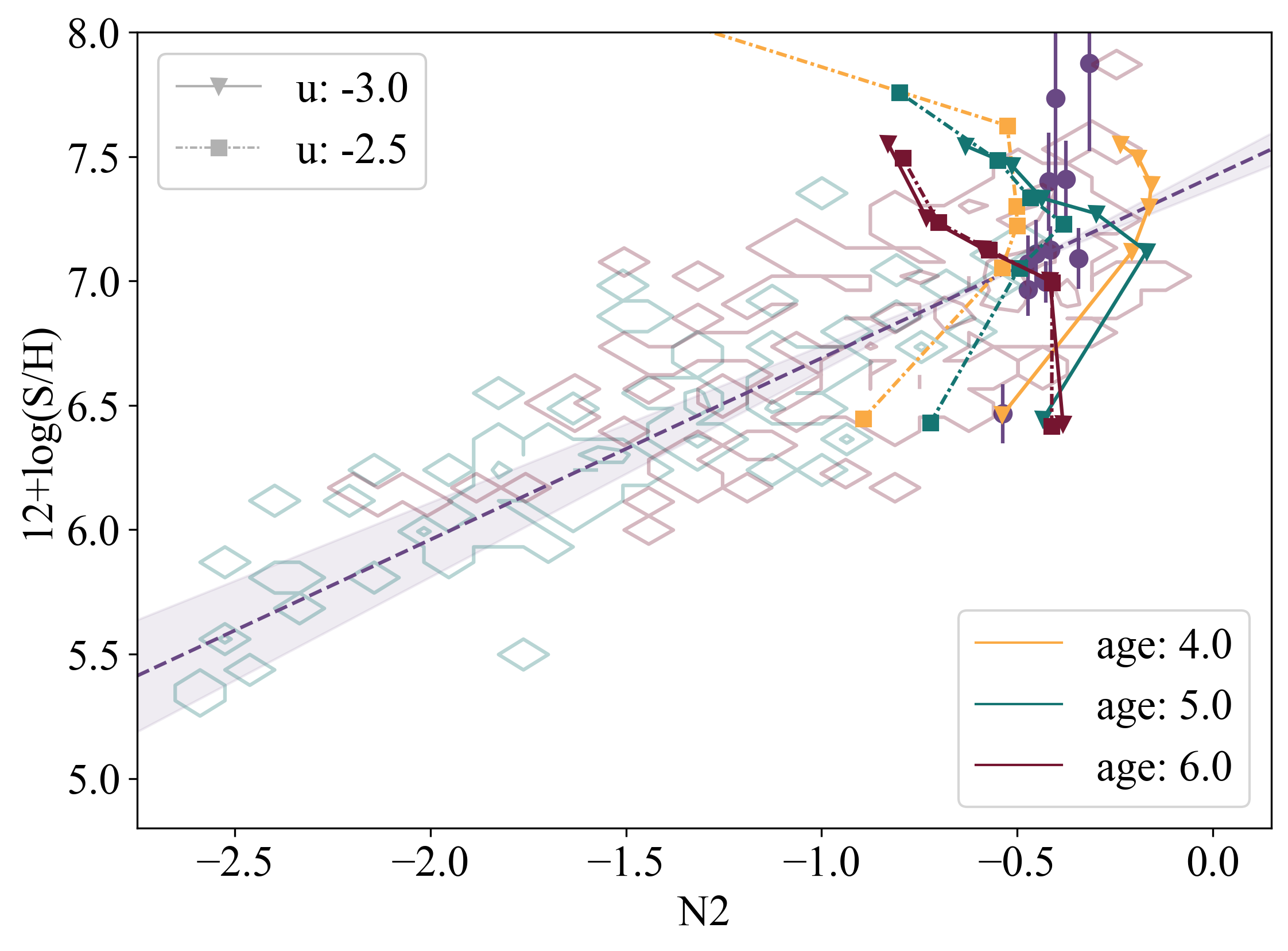}
\caption{The N2 abundance calibration with the sulphur abundance. Red contours correspond to disc HII regions while blue contours correspond to HII galaxies. Purple dots represent the CNFSRs analysed in this work. The Cloudy models described in the text appear superimposed. The calibration from \citet{Denicolo2002} is also show, assuming log(S/O)$_\odot$ $\simeq$ -1.7 \citep{2009ARA&A..47..481A}.} 
 \label{fig:calibracionN2}
\end{figure}

Unfortunately, our wavelength coverage does not include the [OII] emission line, not allowing a direct comparison with their results. However, we can exploit the commonly used N2 parameter, defined as log([NII]$\lambda \lambda$6584/H$\alpha$)  \citep{Denicolo2002} , to obtain an additional independent abundance determination from our data. Then, we used the disc HII regions and HII galaxies presented in \citep{2022MNRAS.511.4377D} to produce a calibration between N2 and the sulphur abundance. Figure \ref{fig:calibracionN2} shows the empirical N2 calibration from \citet{Denicolo2002}, shown as a dashed purple line, scaled under the assumption of a constant S/O solar ratio \citep[$\simeq -1.7$,][]{2009ARA&A..47..481A}. For comparison, we also include the disc HII regions, HII galaxies and the Cloudy models used in Fig. \ref{fig:calibracion}, adopting the same color coding and symbols. It is evident that the calibration also remains valid when sulphur is used as abundance tracer. 

It is important also to remark that the N2 parameter exhibits a large dispersion at high abundances. In contrast to the sulphur calibration, this dispersion appears to be driven not only by the effective temperature of the ionising stellar population but also by variations in the ionisation parameter. Consequently, the turnover observed at these metallicities cannot be easily corrected, introducing an additional source of uncertainty in metal rich regimes. This index has been shown to follow a linear relation with log(O/H) at least over the range 7.2 < 12+log(O/H) < 9.1 \citep{Denicolo2002} which correspond with an upper limit of log(S/H)$\sim$ 7.4. Then, as in the case of the sulphur measurements, our regions we are underestimating the abundances in the regions where the auroral line was not detected. 

The sulphur abundances derived with the direct method (purple dots) show a good agreement with the predictions of the photoionisation models reinforcing the reliability of our measurement. Interestingly, the regions with the highest abundances are well reproduced by models that assume younger stellar populations (4 Myr), as in the case of the S$_{23}$ calibration. In particular, they lie in the region of models with the lower ionisation parameters (log($u$) $\sim -3$). 

Several studies in the literature have investigated the stellar populations of this nuclear ring. Particularly, \citet{Gadotti2019}, using the same dataset analysed in the present work, reported that the young nuclear ring in NGC 1097 exhibits the lowest metal content in the central part of the galaxy \citep[< 50 arcsec, see their Fig. 5 and also Fig. 4 in][]{Bittner2020}. This result is not compatible with the metallicity values derived from the ionised gas, both in this work and in previous studies cited during the discussion. The authors report stellar metallicities as low as [M/H] $\sim$ –1.57 ($\sim$ 3\% solar), a value incompatible with the oxygen and sulphur gas measurements that consistently indicate supersolar abundances. Since the ionising stars (<10 Myr) form directly from this gas, their metallicities cannot be significantly lower than solar.

However, a recent study by \citet{Sextl2025} found that the young stars in this star-forming ring, as well as in other similar ones (NGC 7552, NGC 613, and NGC 335), exhibit supersolar stellar abundances. They performed a pixel-by-pixel SED fitting using the same dataset employed in the previous work, adopting the MILES stellar library \citep{Sanchez-Blazquez2006}, MESA stellar evolution isochrones \citep{Dotter2016,Choi2016}, and a Chabrier initial mass function \citep{Chabrier2003}. However, they realised that the commonly used empirical MILES library includes only stars with temperatures above 9000 K, which is not adequate to describe the complex environments of star-forming nuclear rings. To address this, they extended MILES with additional stellar libraries for hot and post-main-sequence stars \citep[see][]{Sextl2024,Sextl2025}, incorporating hot massive stars, Wolf–Rayet stars, AGB and post-AGB stars, and carbon stars \citep{Eldridge2017,Lancon2000,Aringer2009,Rauch2003,Smith2002}. This improved approach, better suited to the particular characteristics of CNSFRs, revealed metallicities two to three times higher than solar, fully consistent with our results.

\subsection{Ionising cluster characteristics}\label{sec:characteristics}

The number of hydrogen-ionising photons (Q(H$_0$)) in each HII region was calculated from their extinction-corrected H$\alpha$ luminosities, assuming a distance of 14.5 Mpc (see Table \ref{tab:galaxy characteristics}). The related equation was derived using the recombination coefficient of the H$\alpha$ line assuming a constant value of electron density of 100 cm$^{-3}$, a temperature of 10$^4$ K and case B recombination \citep{Osterbrock2006}.

\begin{figure}
\centering
\includegraphics[width=\linewidth]{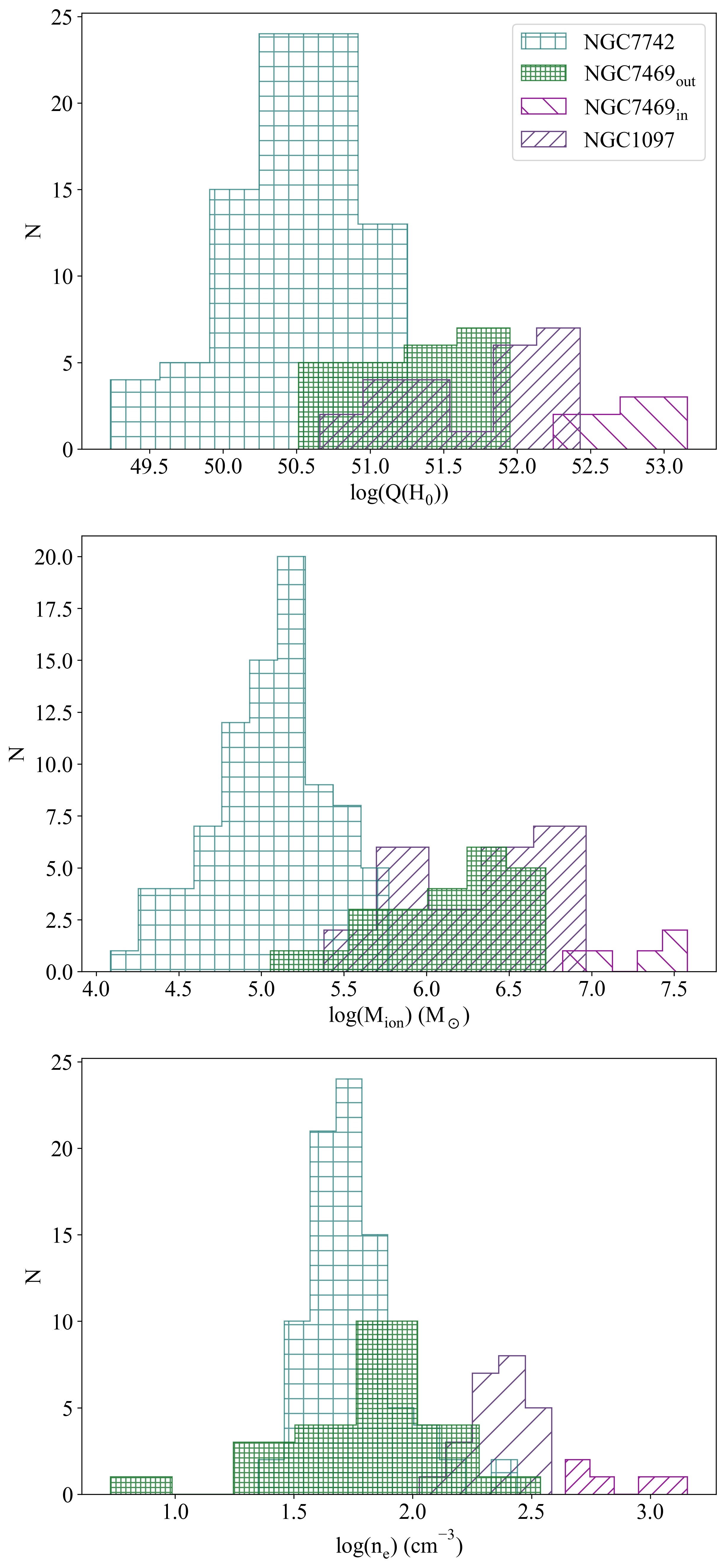}
\caption{The different histograms in the figure show for the ring HII regions the distributions of: the number of hydrogen ionising photons, the ionising mass of the clusters, and  the electron density, upper middle and lower panels respectively.}
\label{fig:hist_r_ne_Q}
\end{figure}

The HII regions show logarithmic H$\alpha$ luminosities ranging from 38.79 to 40.56, with a median value of log L(H$\alpha$) = 40.04. This corresponds to a logarithmic number of ionising photons between 50.66 and 52.43, with a mean value of log(Q(H$_0$)) = 51.90. The upper panel of Fig. \ref{fig:hist_r_ne_Q} shows the distribution of these results. These luminosities are comparable to those observed in the outer ring of NGC 7469 and to the most luminous CNSFRs identified in NGC 7742, despite the closer distance to NGC 1097. However, the regions in the inner ring of NGC 7469 remain about an order of magnitude more luminous.

We used the relation proposed by \citet[][see Equation 30]{ngc7742paper} between the EW(H$\beta$) and the number of ionising photons to estimate the ionising stellar masses of our circumnuclear HII regions. The H$\beta$ equivalent widths range from 5.8 to 19.5 \AA , corresponding to ages between 5.56 Myr and 6.86 Myr for a single burst of star formation, based on PopStar models \citep{Popstar} assuming a Salpeter initial mass function \citep[IMF;][]{Salpeter1955} with lower and upper mass limits of 0.85 and 120 M$_\odot$, respectively. The derived ionising masses range from 2.40 $\times$ 10$^5$ M$_\odot$ to 9.27 $\times$ 10$^6$ M$_\odot$, with a median value of 2.79 $\times$ 10$^6$ M$_\odot$ (see middle panel of Fig. \ref{fig:hist_r_ne_Q}). Then, the total ionising stellar mass in the entire ring is 7.74 $\times$ 10$^7$ M$_\odot$. 

All our regions have masses larger than 10$^4$ M$_\odot$, hence the IFM is assumed to be fully sampled \citep{1994ApJS...91..553G,2010A&A...522A..49V}. Our derived values are consistent with those obtained by \citet{Diaz2007} for the CNSFRs in NGC 2903, NGC 3351, and NGC 3504, higher than those measured in NGC 7742, and lower than those found in NGC 7469. However, these results should be considered as lower limits to the ionising masses, since we assume that (i) there is not dust photon  absorption and re-emission at infrared wavelengths, and (ii) there is not ionising photon escape from the HII regions. 

The electron density was derived from the [SII]$\lambda$6717/[SII]$\lambda$6731 ratio, which is sensitive to densities above n$_e$ = 50 cm$^{-3}$, using PyNeb \citep{pyneb}. Since this ratio shows a slight dependence on temperature, we adopted the mean T$_e$(SIII) value of 5.486 K, obtained from regions with direct electron temperature measurements. All regions in our sample show densities above this lower limit, ranging from 107 $\pm$ 29 cm$^{-3}$ to 385 $\pm$ 37 cm$^{-3}$, with a mean value of 243 cm$^{-3}$. The lower panel of Fig. \ref{fig:hist_r_ne_Q} shows these results. However, although higher than in those found in other CNSFRs, these values remain in the lower limit for densities derived using the [SII] diagnostic.

An important parameter to characterise HII regions is the ionisation parameter, u, which serves as a proxy for the velocity of their ionisation fronts. The dimensionless ionisation parameter is commonly derived from the [SII]/[SIII] ratio. We obtained a range of values -0.061 < log([SII]/[SIII]) < 0.589 in the circumnuclear regions analysed in this work. Other high metallicity CNSFRs reported in the literature also exhibit large [SII]/[SIII] ratios (log([SII]/[SIII]) $\sim$ 0.1 – 0.6), a behavior not observed in typical high metallicity HII regions \citep{Diaz2007}. The authors suggest that the temperature of the ionising source could explain this behavior, with CNSFRs exhibiting a harder radiation field compared to other high-metallicity HII regions. They point out these CNSFR could be affected by hard radiation coming from a low luminosity AGN due to their proximity to the galactic nuclei. However this is not the case of the presented galaxy.

Using the relation proposed by  \citet{Diaz1991}, we obtained ionisation parameters from -3.98 to -2.89, with a median value of -3.42. However, as shown in Section \ref{sec:nature} (see Fig. \ref{fig:bpt}), these high metallicity regions exhibit a noticeable deficiency in the [SII] $\lambda\lambda$ 6717, 6731 \AA\ emission lines. Therefore a revision of this diagnostic is required. To test this issue, we compared the ionisation parameter values with quantities that can be measured independently of the [SII] emission lines. In particular, we compared the physical sizes of the HII regions, derived directly from the segmentation performed on the H$\alpha$ flux map (see Section~\ref{sec:segmentation}), with those estimated from the definition of the ionisation parameter.

\begin{figure}
\includegraphics[width=\columnwidth]{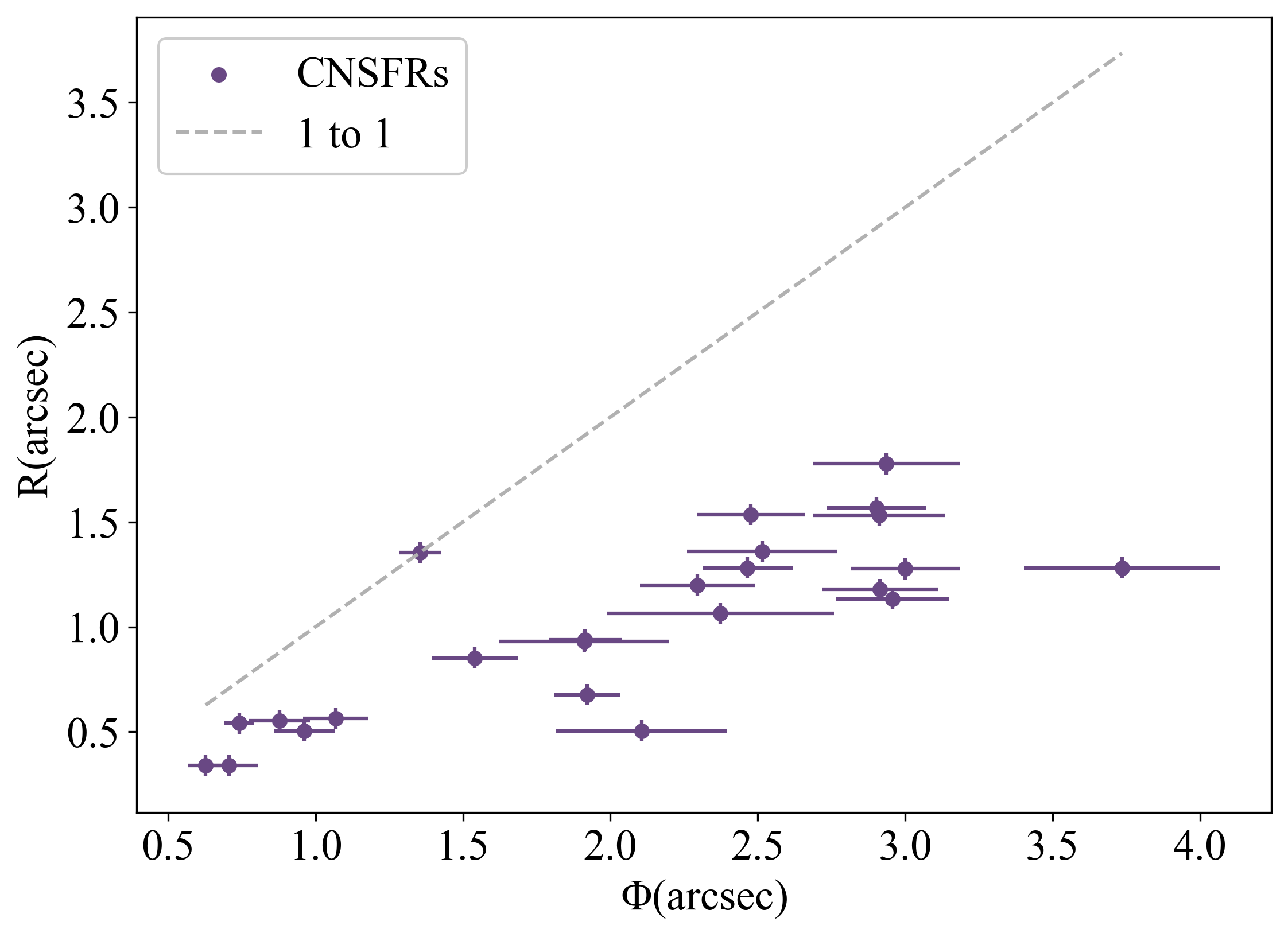}
\caption{The ionisation derived angular radius against the angular radius measured from the HII region segmentation (see Sec. \ref{sec:segmentation}).}
\label{fig:radios}
\end{figure}

We estimated the angular radii of the observed ring HII regions, $\phi$, using the H$\alpha$ fluxes, the electron density, and the ionisation parameter calculated from the [SII]/[SIII] ratio \citep[see][]{2002MNRAS.337..540C,ngc7742paper}. Figure \ref{fig:radios} shows the comparison between the angular radii derived from the ionisation parameter and those measured directly from the H$\alpha$ segmentation. The regions do not lie along the one to one relation expected for ionisation bounded HII regions. Instead, the measured angular radii are systematically smaller than those predicted from the ionisation parameter, with the HII regions following a clear linear trend. Moreover, this behavior cannot be explained by ionising photon escape since, in that case, the HII regions would lie above the one-to-one relation, instead of below it.

Only one region, R1, lies within the area expected for ionisation bounded nebulae. However, in the H$\alpha$ emission map, we detect a possible overlap between this ionised region and a nearby one, which could artificially increase the measured radius without significantly affecting the H$\alpha$ flux. Additionally, we tried to force the segmentation procedure to select larger HII regions by decreasing the background threshold in the H$\alpha$ emission map. However, this new procedure did not resolve the discrepancy. Instead, it shifted the regions diagonally toward the upper right part of the plot.

In previous studies of other circumnuclear rings in this series of papers, we observed different behaviors of their HII regions. In NGC7742, the CNSFRs are ionisation bounded, with all measured radii consistent with those estimated from the ionisation parameter, $\phi$ \citep{ngc7742paper}. In contrast, the CNSFRs in both rings of NGC 7469 appear consistent with bubbles inflated by stellar winds originating from Wolf–Rayet stars \citep{NGC7469paper}. 
The main difference between NGC 1097 and these two example galaxies is the higher density in its CNSFRs, which apparently is related to the [SII]/[SIII] ratio.

\begin{figure}
\centering
\includegraphics[width=\linewidth]{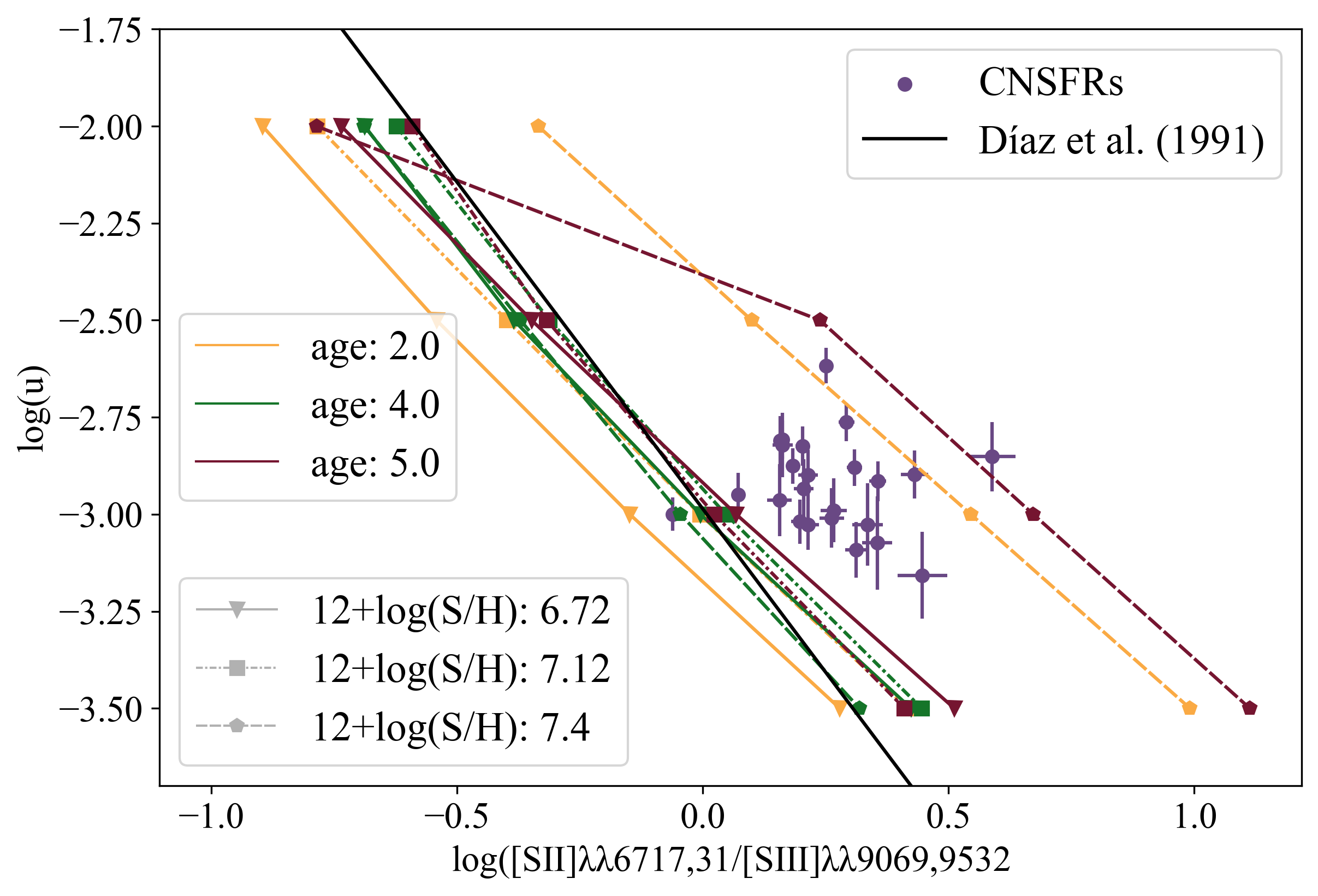}
\caption{The ionisation parameter derived using the angular radius measured from the HII region segmentation against the logarithmic [SII]/[SIII] emmision ratio.}
\label{fig:new_u}
\end{figure}

We calculated the ionisation parameter using its definition, u = Q(H$_0$)/($4\pi c n_e R^2$), obtaining values between –3.045 and –2.505 in logarithmic scale. These values are systematically higher than those derived previously using the empirical calibration by \citet{Diaz1991}. Figure \ref{fig:new_u} shows the ionisation parameter as a function of the [SII]/[SIII] emission-line ratio, with the empirical calibration previously used for comparison. In order to better understand the relation between the [SII]/[SIII] ratio at high metallicities, we used the CLOUDY photoionisation models described in Section \ref{sec:supersolar}. To explore the dependence on the effective temperature of the ionising source, we selected stellar cluster ages of 2, 4, and 6 Myr, representing populations both without and with significant Wolf–Rayet (WR) emission. In the models implemented \citep{Popstar}, the WR stars appear around 4 Myr. We adopted four values for the ionisation parameter, log(u) = –3.5, –3.0, –2.5, and –2.0, and three metallicities, 12 + log(S/H) = 6.72, 7.12, and 7.40 (corresponding to 0.4, 1, and 2 times the solar value respectively), assuming that the stellar metallicity is equal to that of the gas. These photoionisation models are also overplotted in Figure \ref{fig:new_u}.

As anticipated, the HII regions analysed in this work are in the right part of the log(u)–[SII]/[SIII] calibration, implying that the ionisation parameter is systematically underestimated when derived it from this ratio. From the models, we observe a change in the trend for metallicities above solar, coinciding with the turnover point of the S$_{23}$ calibration. While most models follow the relation established by \citet{Diaz1991}, those with supersolar metallicities and without WR stars (2 and 5 Myr), and thus with stellar clusters with lower effective temperatures, exhibit higher [SII]/[SIII] ratios. The models with supersolar abundances and WR stars (at 4 Myr) appear to be consistent with the ionisation parameter calibration. Therefore, a change in the ionisation structure of HII regions with supersolar metallicities is the most plausible explanation for their elevated [SII]/[SIII] ratios.

Notably, this behavior is not exclusive to the [SII]/[SIII] ratio. The commonly used [OII]/[OIII] ratio also increases with metallicity, although this trend appears at lower abundances \citep[see Figure 18 of][]{Diaz2007}. Therefore, a detailled study of the ionisation structure could provide valuable insights into the properties of these high metallicity ionising sources. In the following section, we explore alternative explanations related specifically to the ionisation conditions and structure (see Section \ref{sec:temperature-bounded}).

Finally, the mass of ionised hydrogen, in solar masses, have be derived using the expression given in \citep{Diaz1991} and the ionisation parameter calculated from the definition. Their median mass is 1.1 $\times$ 10$^5$ M$_\odot$, covering a range of values between 6.8 $\times$ 10$^3$ M$_\odot$ and 2.7 $\times$ 10$^5$ M$_\odot$. The total mass of ionised hydrogen in the ring is $\sim$ 2.6 $\times$ 10$^6$ M$_\odot$.

Table \ref{tab:HIIcharacteristics} (see Appendix \ref{ap:pop}) shows the characteristics of each HII region and lists in column 1 to 7: (1) the region ID; (2) the extinction-corrected H$\alpha$ luminosity; (3) the number of hydrogen ionising photons; (4) the ionisation parameter estimated from the definition; (5) the electron density; (6) the mass of ionised hydrogen and (7) the ionising mass.

\subsection{Temperature bounded emission}\label{sec:temperature-bounded}

The empirical relation introduced in \citep{Diaz1991} between the ionisation parameter and the [SII]/[SIII] emission line ratio upholds for a wide range of observations. However, we present the first evidence, as far as we know, where the trend breaks, as seen in Figure \ref{fig:new_u}. Since we compute the ionisation parameter  using its definition, the further analysis would follow the [SII]/[SIII] shift in behavior at high metallicity. Although this emission line ratio is independent from abundance growth by definition, an increase in metal content could produce significant changes in the HII region structure and collisional exited line emissivities.

A more metallic young stellar population presents a softer ionising spectrum, which imposes a different ionisation structure once the recombination equilibrium is established. Similar to the Q(H$_0$) parameter, one can define the Q(S$_1$) as the number of ionising photons respect to the single ionised sulfur, whose ionszation potential is 23.34 eV. A decrement in Q(S$_1$) leads to a shortening of the [SIII] emission part of the nebula, and therefore, produce an enlargement of the [SII] dominated region, which always extends to the very end of the nebula since the neutral sulfur ionisation potential (10.36 eV) is lower than the hydrogen one. This change in the recombination equilibrium between the two sulfur ionic species results in a higher [SII]/[SIII] ratio.

\begin{figure*}
\centering
\includegraphics[width=0.44\linewidth]{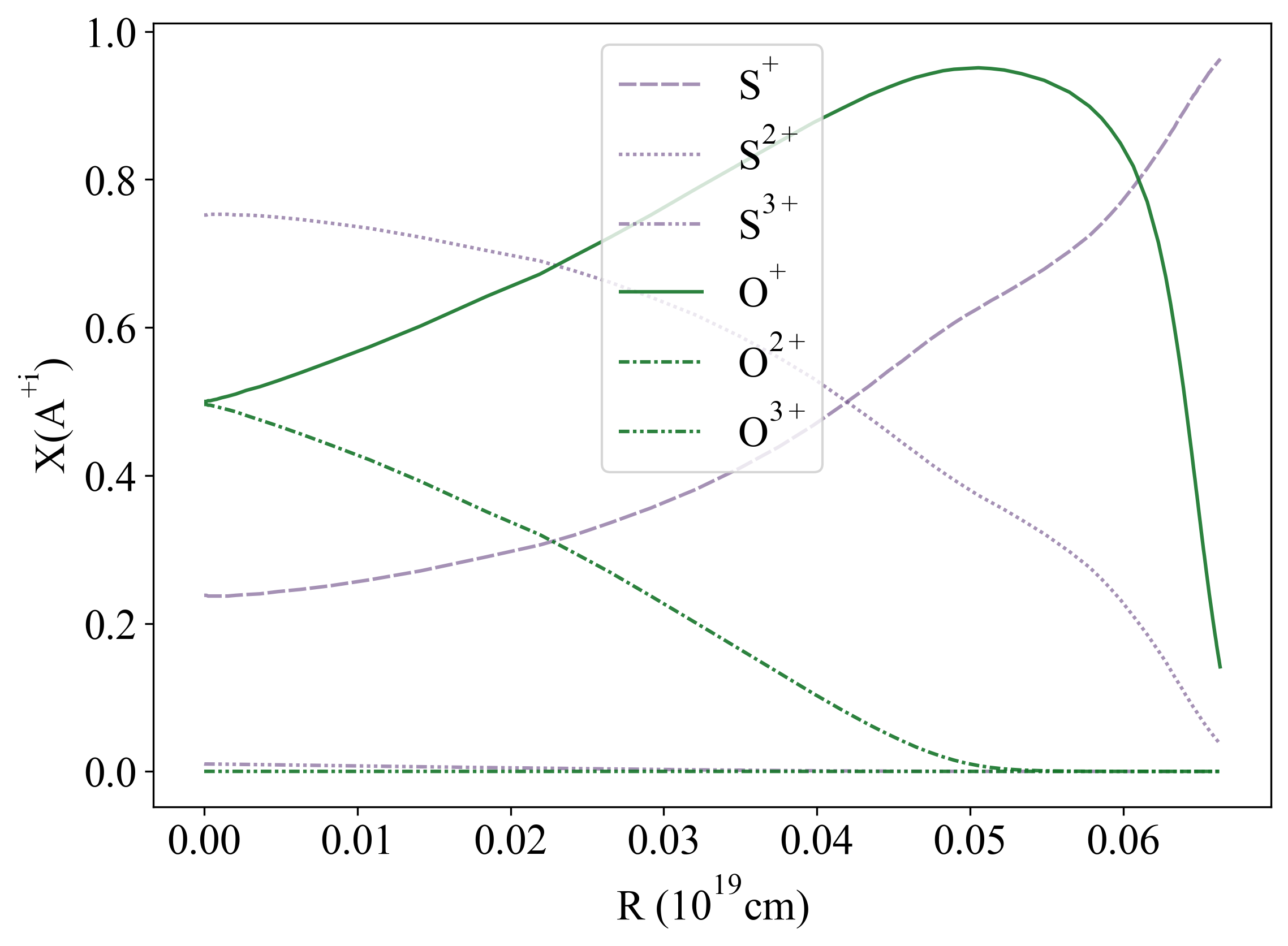}
\includegraphics[width=0.49\linewidth]{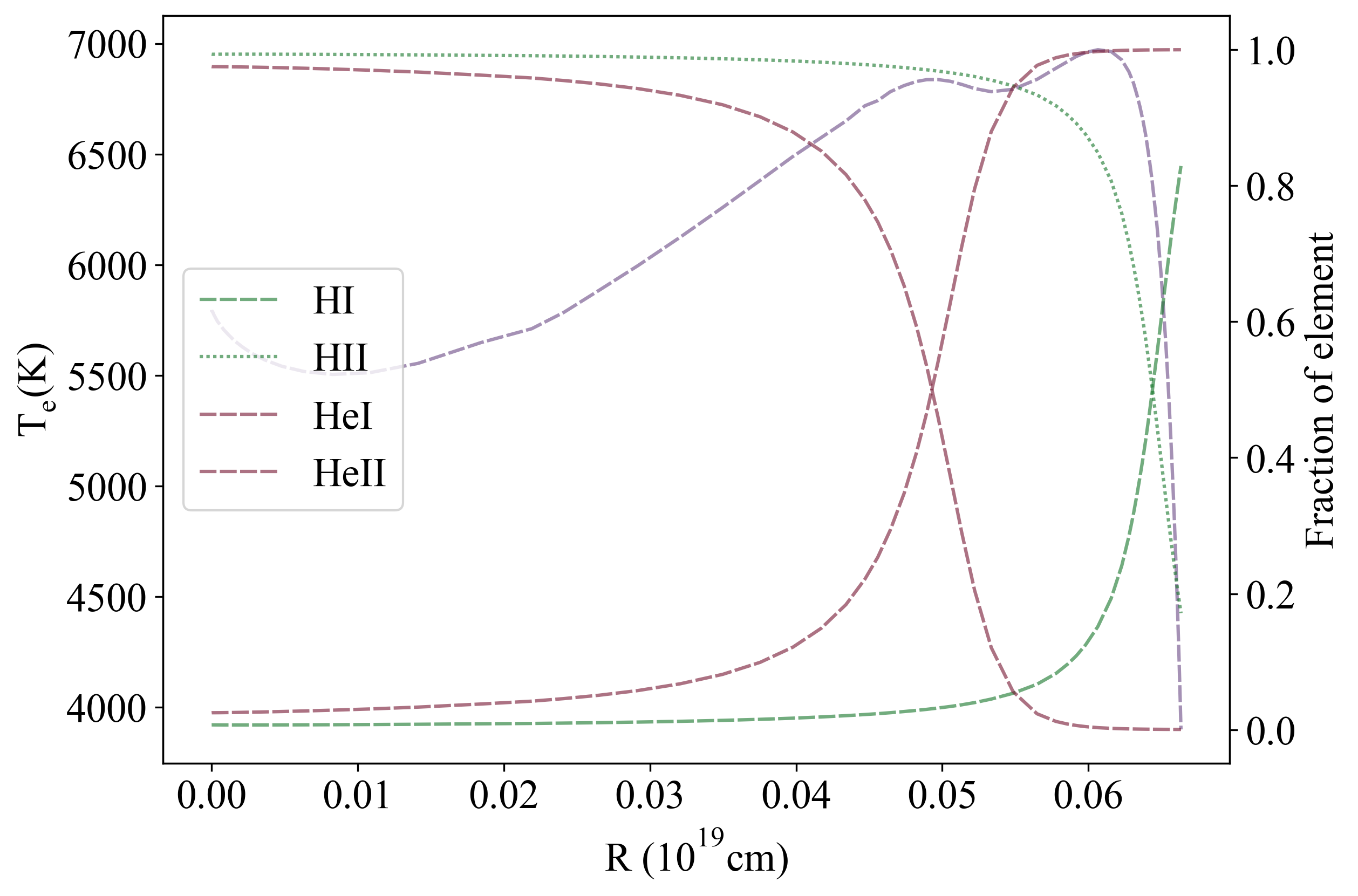}
\includegraphics[width=0.44\linewidth]{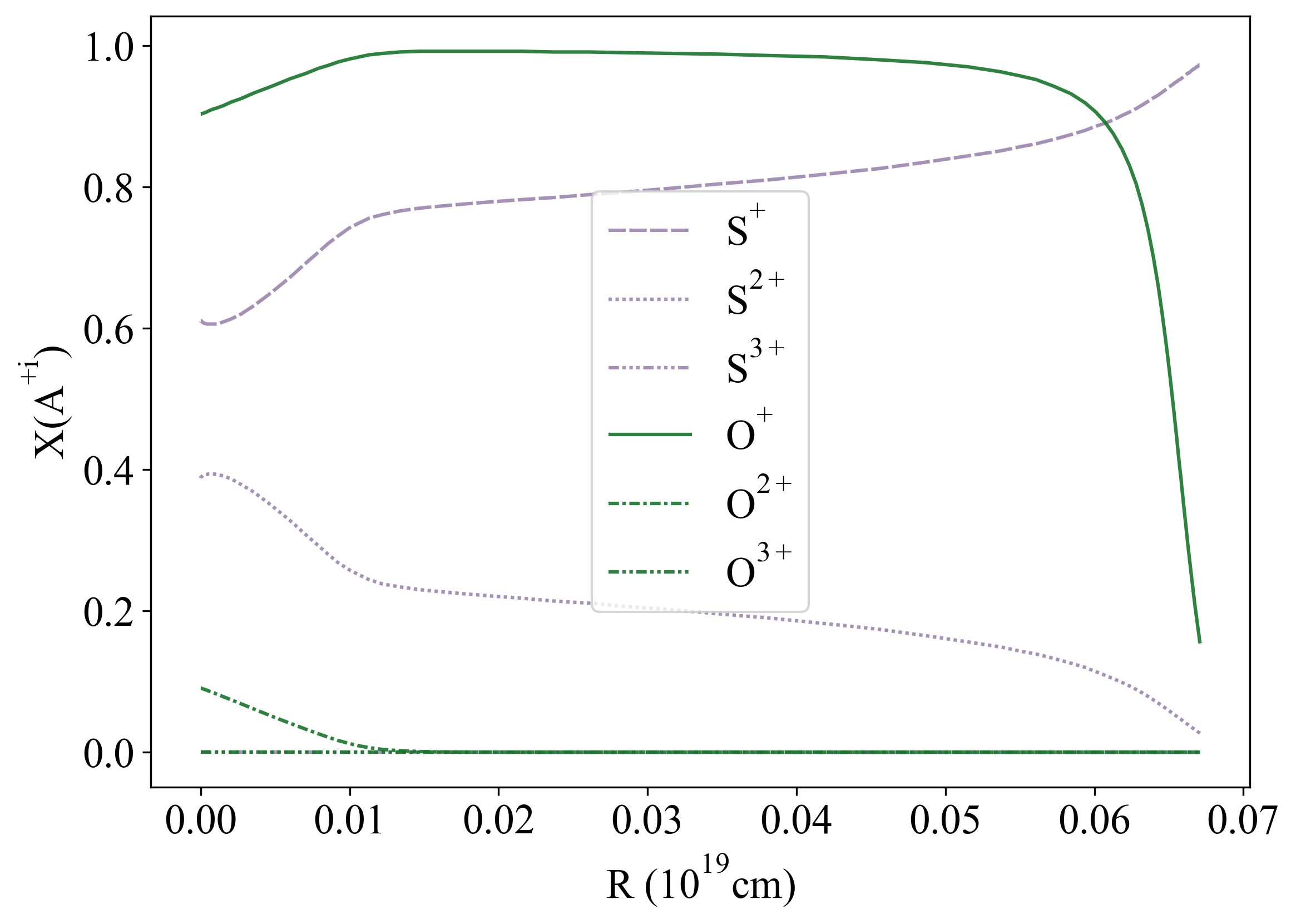}
\includegraphics[width=0.49\linewidth]{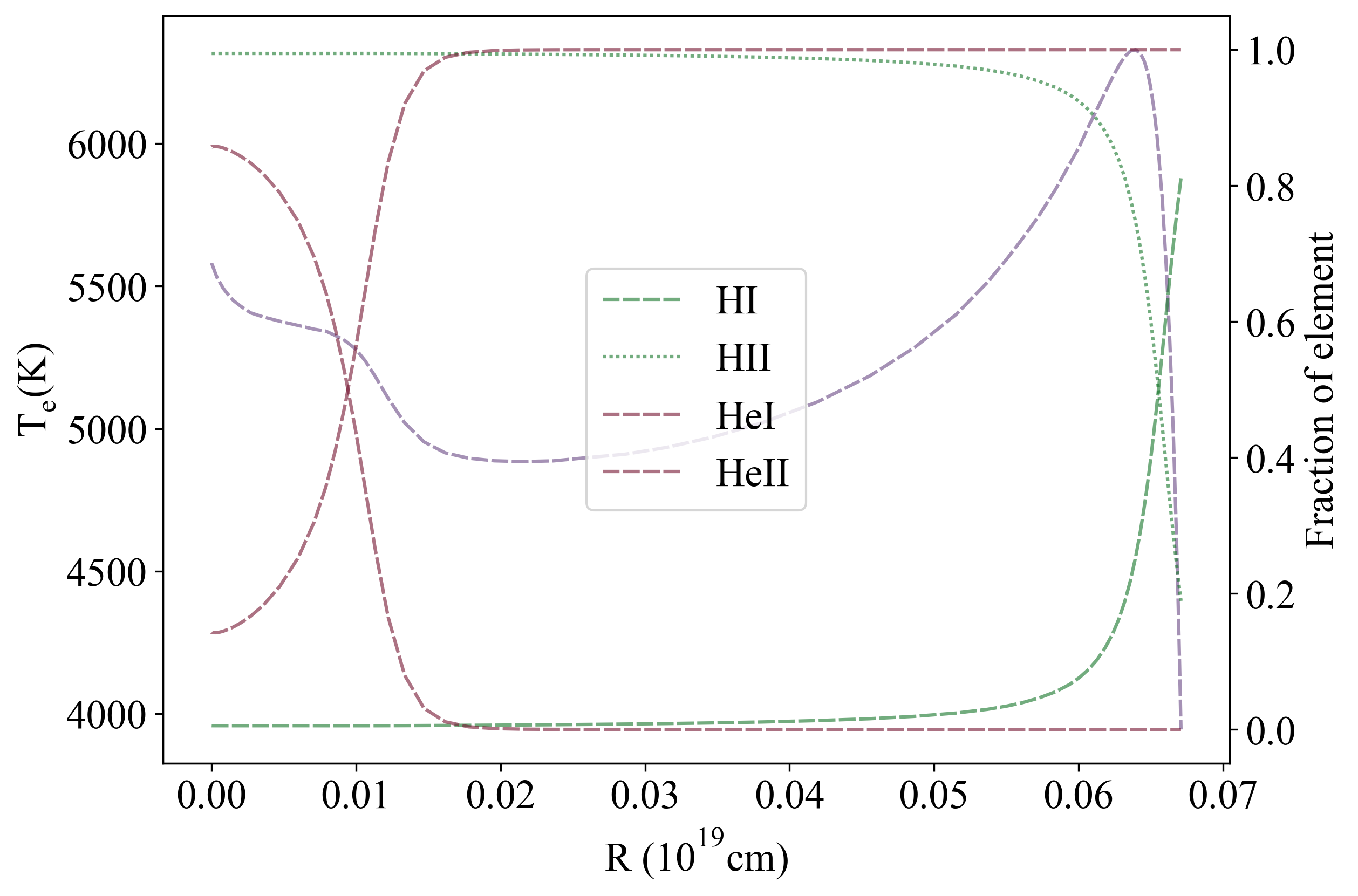}
 \caption{Ionisation structures of the same Cloudy model presented in Figure \ref{fig:new_u}. Their metallicity is log(S/H) = 7.4, the ionisation parameter is similar to the one found in our CNSFRs (log(u) = 3), and ages are 2 and 5 Myr, top and lower panels respectively. In the right panels appear the oxigen and sulphur structures while in those in the right we presented the hydrogen and helium toguether with the electron temperature thought the nebula. }
 \label{fig:ion_structure}
\end{figure*}

As we can see in Figure \ref{fig:ion_structure}, the older the stellar population, the lower the Q(S$_1$) and the bigger the low ionisation region became. Nevertheless, this trend is broken by the stellar population of 4 Myr old, since at this stage the spectrum is dominated by wolf-Rayet stars, whose effective temperature are much higher, setting the [SII]/[SIII] ratio back to the \citep{Diaz1991} relation.

Nonetheless, the fact that the two ionisation structures shown in Figure \ref{fig:ion_structure} generate a similar [SII]/[SIII] ratio is a truly surprising result. Whose interpretation requires a deeper understanding of the involved emission line formation processes.

Collisional de-excitation lines are forbidden atomic transitions triggered by the interaction of the ion with a free electron, resulting in the population of an excited state. Which, in the absence of further collisions, decays to a lower energy level emitting a photon. Additionally, free electrons in an HII region follow an Maxwell-Boltzmann distribution, characterized by the electron temperature (Te), being the only thermalized species in the nebula. And therefore, Te sets the collisional excitation rate for the rest of ionic species.

In general, the relation between the temperature of the electron distribution and the energy of the atomic transition can operate under three regimes. On the one hand, when the peak of the Maxwell-Boltzmann distribution exceeds the transition energy ($1/2KTe > E_{cel}$) the collisional excitation rate is low, since the transition is transparent for most of the electrons due to their high energy. On the other hand, when the electron temperature equals the transition energy ($KTe\eqsim E_{cel}$) the collisional excitation rate reaches its maximum. Finally, when the electron distribution further shifts to lower temperatures the collisional excitation rate decrees, since a growing number of electrons don't have enough energy to excited the upper transition level; causing a temperature bounded emission regime. In the case of the [SII] $^{2}{D}$ - $^{4}{S}$ transition ($\lambda$6717 and $\lambda$6731 lines) the transition energy correspond to 1.85 eV, from where we can divide the different boundaries of the three regimes approximately at $Te\gtrsim43000K$ and $Te\lesssim5800K$. The former correspond to the Te when the mode of the electron distribution equals the transition energy; while the later is define by the transition energy appearing more than 2 standard deviations away from the mean electron temperature ($3/2 KTe -2\cdot \sqrt{3/2K^{2}Te^{2}} < E_{cel}$).

At the high metallicity regime the metal cooling is higher and the electron temperature is low ($\sim 5000K$), thus the nebula operates in the later mentioned regime  ($3/2 KTe< E_{cel}$). Therefore, even though the $S^{+}$ ion dominate the ionisation structure their emissivity is limited by the T$_e$. There are just a few electrons from Maxwell-Boltzmann distribution that can excite the [SII] $^{2}{D}$ - $^{4}{S}$ transition in the low excitation part of the nebulae. This low collisional excitation rate explains the lesser increase in the [SII]/[SIII] ratio seen in Figure \ref{fig:new_u}, as well as why the observations fall short in the [SII]/H$\alpha$ ratio with respect to the models of Figure \ref{fig:bpt}. Nonetheless, the effects of low electron temperature upon the collisional excitation line emission will benefit of further exploration in future works.

\subsection{Comparison with previous studies}

In this section, we compare our results with those from the most detailed photometric study to date, performed by \citet{Prieto2019}, who used infrared-to-ultraviolet imaging with a spatial resolution of $\sim$ 10 pc. We also contrast our results with those obtained for the 14 regions selected in Br$\gamma$ emission (IR) by \citet{Kotilainen2000}, and the 10 regions identified using radio continuum emission at 1.465 GHz by \citet{Hummel1987} (see Section \ref{sec:segmentation} for details). Table \ref{tab:multiw} presents the identification of our regions with those reported by the comparison studies, along with some of the key properties discussed throughout this section.

In \citet{Prieto2019}, they identified 247 individual star clusters within the ring with sizes of FWHM < 8 pc; 171 were detected in the UV and 76 at 0.8 $\mu$m. Assuming a density of 100 cm$^{-3}$, a radius of 8 pc implies a number of ionising photons of log(Q(H$_0$)) = 50.2, similar to that produced by single O4 – O5 stars \citep[see][]{Sternberg2003}, while the number of ionising photons we measure is up to one order of magnitude higher. Therefore, when comparing our results with those of \citet{Prieto2019}, it is important to note that their analysis primarily traces regions that may be ionised by individual massive stars, rather than by ionising clusters as in the present work. Moreover, while \citet{Prieto2019} directly analyzed the stellar emission, our approach infers their physical properties indirectly through the analysis of the surrounding ionised gas emission. Due to the extremely high spatial resolution of their observations and the inherently different methodologies applied, we focused our comparison in the global properties of the regions in the ring, instead of a one by one cluster comparison.

The excitation mechanism of the hot molecular gas can be observed using the H$_2$/Br$\gamma$ ratios. In \citet[][]{Kotilainen2000} (see Tab. \ref{tab:multiw}), this ratio ranges from 0.25 to 1.70, with an average value of 0.75, which can be explained by UV excitation from young massive stars \citep[see][]{Puxley1990}. This result is consistent with our, where photoionisation processes successfully reproduce the [OIII]/[NII] emission ratios observed in the CNSFRs of the ring. Only two regions, K2 and K9, exhibit H$_2$/Br$\gamma$ ratios higher than 0.9. They are located at the opposite ends of the inner galaxy bar, and notably, they are not detected in the optical Balmer emission as shown in the present work. Also, the regions with higher H$_2$/Br$\gamma$ ratio are the ones with lower radio emission. A possible explanation is a higher gas density in them or an additional contribution from shock excitation, likely driven by supernova activity. This scenario is consistent with the idea of a propagating starburst that originates in the nucleus and extends toward the ends of the bar and into the circumnuclear ring through bar driven shocks. We  found just a few  clusters in the filaments that spiral to the centre or in the inner bulge, and none along the filaments outside the ring.

The dust extinction of the regions in the ring has been estimated by several authors using different indicators. From photometric studies, \citet{Barth1995} found a mean attenuation of A$_V$ = 1.1 mag using Hubble Space Telescope images. In \citet{Prieto2019}, they derived an average extinction for the cluster continuum of A$_V$ < 1.5 mag, consistent with the very blue spectral slopes of the clusters. In contrast, the extinction associated with the interstellar medium is found to be slightly higher, with A$_V$ $\sim$ 2 – 3 mag, as inferred from extinction maps of the circumnuclear ring on kiloparsec scales. This is consistent with the extinction measured for the CNSFRs in this work (A$_V$ = 2.14 mag) and with the values derived from the Pa$\alpha$/H$\alpha$ ratio reported by \citet{Prieto2019}. Using H$\alpha$/H$\beta$ emission, \citet{Hummel1987} reported 0.85 mag and 2.83 mag for the eastern and western parts of the ring respectively, \citet{Osmer1974} measured 0.56 mag for a northern region, \citet{Phillips1984} found 3.0 and 3.5 mag in two other regions, and \citet{Walsh1986} reported a range between 0.6 and 3.0 mag. Finally, \citet{Kotilainen2000} used the Br$\gamma$ fluxes, obtained A$_V$ values between 0.1 and 2.2 mag, with an average of 1.3 mag. These measurements are broadly consistent with our results.

The ionised gas masses of the regions reported by \citet{Kotilainen2000} range from (1.7 – 8.5) $\times$ 10$^6$ M$_\odot$, summing to a total of 5.1 $\times$ 10$^7$ M$_\odot$ for the ring, comparable to the value we derived, 7.74 $\times$ 10$^7$ M$_\odot$. In contrast, \citet{Prieto2019} estimated a lower total stellar mass formed in the ring over the past 20 – 100 Myr, of approximately $8 \times 10^{6}$ M$_\odot$. The mass associated with their young stellar population is $\sim 3 \times 10^{6}$ M$_\odot$, distributed among roughly 60 clusters with a mean mass of $\sim 5 \times 10^{5}$ M$_\odot$ per cluster (see their Fig. 4). If the mass of the additional 76 clusters detected but not analysed in that work is included, the total stellar mass increases to $\sim 1.5 \times 10^{7}$ M$_\odot$. This value is a factor of $\sim$ 2 – 4 lower than we inferred from the ionised gas, suggesting that the ionised gas traces a more efficient conversion of molecular gas into stellar mass. The molecular gas mass in the ring is M(H$2$) $\sim 10^9$ M$_\odot$, derived by \citet{Hsieh2011} from HCN observations, and by \citet{Gerin1988} from CO(J = 1-0). \citet{Prieto2019} found that the total mass of clusters formed represents less than 1\% of this gas, whereas our results show that the ionising stellar mass (< 6 Myr) is $\sim$ 8\% of the gas content of the ring. 

The star formation rate (SFR) was calculated by \citet{Kotilainen2000} by comparing the observed quantities with models from \citet{Leitherer1999} and assuming an instantaneous star formation burst. They found the SFR for the star forming clumps are between 0.05 – 0.31 M$_\odot$ /yr, a total of 1.8 M$_\odot$/yr within the ring. This last value is the same found by \citet{Prieto2019}, calculated using the extinction corrected Pa$\alpha$ emission map of the ring and the \citet{Kennicutt2012} relation, and it is a factor 2.5 lower than previous estimate by \citet[][$\sim$ 5 M$_\odot$]{Hummel1987}. Taking into account the molecular mass gas, it implies a gas depletion time-scale, M(H$_2$)/SFR, of 5 $\times$ 10$^8$ yr. Additionally, with this SFR constant during the past 100 Myr, the total mas in clusters in the ring should be 2 $\times$ 10$^8$ M$_\odot$, which is more than one order of magnitude higher than the stellar mass they found within this age range. They also estimated a gas inflow rate of 3 M$_\odot$/yr into the ring through the two dust lanes, and an additional 0.6 M$_\odot$/yr inflow toward the nucleus through the inner spiral lanes. Then, the estimated gas consumption in the ring is 2.4 M$_\odot$/yr, larger than the current star formation they found in the ring.

On the other hand, we calculated the SFR of the CNSFRs directly from the extinction corrected H$\alpha$ emission using also the \citet{Kennicutt2012} calibration, obtaining a total SFR of 2.34 M$_\odot$/yr for all regions, value similar to the gas consumption estimated for the ring in \citet{Prieto2019}. This suggests that the gas supply within the ring is the main factor regulating its star formation activity. At this rate, the ring would have produced $\sim$1.4 $\times$ 10$^7$ M$_\odot$ of stars over the past 6 Myr. Given the total ionising stellar mass derived, this implies that the recent star formation episode must have experienced an enhancement. We estimate a characteristic gas timescale of $\sim$4 $\times$ 10$^8$ yr for this process.

Finally, \citet{Prieto2019} argue that the HII regions in this circumnuclear star-forming ring are optically thin, with a significant fraction of the ionized gas escaping the boundaries of individual clusters and mixing with the ionized gas. However, this scenario is not supported by the results of this work. From the comparison between the angular radii of the observed HII regions, $\phi$, and those measured directly from the H$\alpha$ segmentation, we show that the properties of our regions cannot be explained by ionizing photon escape (see Sect. \ref{sec:characteristics}). Moreover, the gas we observe cannot originate from leaked ionizing photons, as the measured emission line ratios are inconsistent with those typically associated with such diffuse media (see Sect. \ref{sec:nature}). In particular, the [SII]/H$\alpha$ ratio in our regions has a mean value of 0.16, whereas warm ionized or diffuse ionized gas (WIM or DIG) typically exhibits values greater than 0.4. Similarly, the derived physical conditions as electron densities exceeding 100 cm$^{-3}$ ($n_e$ $\sim$ 240 cm$^{-3}$) and electron temperatures of T$_e$ $\sim$ 5480 K, are incompatible with those expected for diffuse ionized gas, which typically has n$_e$ $\sim$ 0.1 cm$^{-3}$ and T$_e$ $\sim$ 10000 K \citep[see][]{Reynolds2004}. We refer to \citet{Prieto2019} for an alternative discussion of this issue.

\section{Summary and conclusions}
In this work, we studied the ionised gas and the chemical abundances of the CNSFR within the ring of the barred spiral galaxy NGC 1097 using publicly available MUSE IFS observations which cover the optical rest frame part of the spectrum from 4800 to 9300 \AA. Although this ring has been extensively studied both photometrically and spectroscopically over the past decades, using a variety of approaches and instruments across different wavelength ranges, the new IFS data provides a much better spatial resolution at optical wavelengths.

The H$\alpha$ and [OIII] emission line maps reveal two distinct streams that coincide with the location of one of the jets observed in this galaxy. These arc-structures exhibit elevated [OIII]/[NII] ratios, consistent with shocked gas driven by the activity generated by its AGN . The continuum emission is originated from the center of the galaxy, and two armed structures are clearly identified. The bar shows  more prominently in the redder continuum, likely because it is dominated by an older stellar population, whereas the ionising star forming clusters show enhanced emission at bluer wavelengths, reflecting their young stellar populations.

The projected radial extent of the ring, derived from the pixel-by-pixel H$\alpha$ radial profile, spans from approximately 385 pc to 1.3 kpc. Within this structure, we identified a total of 24 HII regions. We extracted the spectrum of each region and measured the fluxes of their main emission lines, including the auroral [SIII]$\lambda$6312 \AA\ line, which was detected in approximately $\sim$ 45\% of the regions. This allowed us to derive direct total sulphur abundances, with a median value close to the solar abundance, 12+log(S/H) = 7.11, and ionic fractions S$^+$/S ranging between $\sim$ 45 - 70\%. The highest measured abundance exceeds five times the solar value, with T$_e$([SIII]) = 3912 $\pm$ 567 K and 12+log(S/H) = 7.88 $\pm$ 0.35, representing, to our knowledge, the highest sulphur abundance reported to date. Although this value may appear extreme, photoionisation models computed for this region successfully reproduce the observed line ratios, being consistent with models that assume young stellar populations (4 Myr) and low ionisation parameters (log(u) $\sim$ -3). In addition, the CNSFR in this galaxy exhibit high oxygen abundances, as previously reported in the literature, with log(O/H) = 9.40 and 9.28 (assuming log(O/H)$_\odot$ = 8.92), as well as stellar metallicities two to three times higher than solar. Under such extreme conditions, commonly used empirical abundance diagnostics, such as $O_{23}$ (commonly refer to as $R_{23}$) and $S_{23}$, are not readily applicable, since the regions lie on the upper branch, out of the range of both calibrations. Similarly, stellar abundance determinations must be treated with caution, requiring the inclusion of stellar populations with effective temperatures below 9000 K, which are the appropriate ones for describing the complex environments of star-forming nuclear rings.

The high metal content in these regions leads to significant changes in the ionisation structure and in the emissivities of collisionally excited lines. These effects alter the recombination equilibrium between the two sulphur ionic species, resulting in enhanced [SII]/[SIII] ratios. A similar behaviour is also observed in the commonly used [OII]/[OIII] ratio, although this trend becomes apparent at lower metallicities.

We compared our results with those reported in the literature. The observed H$_2$/Br$\gamma$ ratios can be naturally explained by UV excitation from young massive stars, in agreement with our results. In particular, photoionisation processes are able to successfully reproduce the [OIII]/[NII] line ratios observed in the CNSFR. The dust extinction of the regions was estimated using different indicators. Values derived from photometric studies and from Br$\gamma$ fluxes are lower than the mean extinction obtained in this work, whereas those inferred from the H$\alpha$/H$\beta$ ratio are broadly consistent with our results (mean value A$_V$ = 2.14 mag). The ionised gas masses derived from the observed infrared emission sum to a total of 5.1 $\times$ 10$^{7}$ M$_{\odot}$ for the ring, which is comparable to the value obtained in this work, 7.74 $\times$ 10$^{7}$ M$_{\odot}$. The molecular gas mass in the ring is M(H$_2$) $\sim$ 10$^{9}$ M$_{\odot}$, as inferred from HCN and CO($J=1$--0) observations. Therefore, the mass in ionising stars younger than 6 Myr represents approximately $\sim 8\%$ of the total gas content of the ring. The SFR reported by other authors for the individual star forming clumps range between 0.05 – 0.31 M$_\odot$/yr, yielding a total SFR of 1.8 M$_\odot$/yr within the ring. This value was obtained both by comparing the observed infrared quantities with model predictions and by using the extinction-corrected Pa$\alpha$ emission map of the ring. However, the SFR directly derived in this work from the extinction-corrected H$\alpha$ emission, yields a higher value, with a total SFR of 2.34 M$_\odot$/yr for all regions. Our estimate is fully consistent with the expected gas consumption in the ring, given the gas inflow rate through the two dust lanes (3 M$_\odot$/yr) against the subsequent inflow toward the nucleus through the inner spiral lanes (0.6 M$_\odot$/yr).

In the following, we present our main conclusions and an overall picture of the circumnuclear ring:
\begin{itemize}
\item Despite the activity and variability of the galaxy nucleus,  no shock effects are observed in the CNSFR. This results in a complex, multi-epoch interaction between the core and the ring, while ensuring that the H II regions are purely photoionised and hence the calculations performed in this work are valid.
\item The directly derived abundances reach four and five times the solar value, and photoionisation models computed for these regions successfully reproduce the observed line ratios. This extremely high-metallicity regime is present, but it has likely not been extensively studied because these regions lie out of the calibration range of commonly used abundance diagnostics, such as O$_{23}$ and S$_{23}$.
\item At these high abundances, we find a break in the ionisation-parameter–[SII]/[SIII] relation that cannot be explained by photon escape. However, changes in the ionisation structure and line emissivities explain the breakdown and the models support this interpretation. Also, the commonly used [OII]/[OIII] ratio shows this behavior although the trend appears at lower metallicities.
\item The starburst is generated in the galaxy nucleus and extends along the inner bar into a ring, generating socks at the ends. Young stars ionise  8\% of the total gas in the ring, and the gas supply regulates its SFR.
\end{itemize}

After this work, the general picture of the circumnuclear ring has significantly evolved. The ring is composed of at least 24 star forming complexes, with the gas supply acting as the primary factor regulating its star formation activity. Multiwavelength studies support the idea of a propagating starburst that originates in the nucleus and extends toward the ends of the bar and into the circumnuclear ring through bar driven shocks, with the regions placed in the opposite ends of the bar not detected in the optical Balmer emission and exhibiting high H$_2$/Br$\gamma$ ratios. The metal content of the ionised regions is extremely high, reaching up to five times the solar value, as expected in the central regions of spiral galaxies. Finally, we likely detect optical signatures associated with one of the two known jets in this galaxy: two arc-structures that are consistent with shocked gas driven by AGN activity. This finding, together with the radio core emission previously found at sub-parsec scales, reflects the presence of feedback processes operating even on small galactic disc scales (< 1 kpc); however, the star formation occurring in the nuclear ring of this specific galaxy does not appear to be directly affected.

\begin{acknowledgements}
We thank A. Prieto for the useful discussion regarding our results in the context of her previous photometric study. This research has made use of the services of the ESO Science Archive Facility and NASA’s Astrophysics Data System Abstract Service. It is based on observations collected at the European Organisation for Astronomical Research in the Southern Hemisphere under ESO programme 097.B-0640(A) and data products created thereof. Also we have used observations obtained with the NASA/ESA HST and obtained from the Hubble Legacy Archive, which is a collaboration between the Space Telescope Science Institute (STScI/NASA), the Space Telescope European Coordinating Facility (ST-ECF/ESA), and the Canadian Astronomy Data Centre (CADC/NRC/CSA). 

This work has been funded by project Estallidos8 PID2022-136598NB-C33 (Spanish Ministerio de Ciencia e Innovacion). 
SZ acknowledges support from the European Union (ERC, WINGS, 101040227). 
\end{acknowledgements}

\bibliographystyle{aa} 
\bibliography{bibliografia}

\newpage
\appendix
\onecolumn
\section{Tables}\label{ap:pop}

\begin{table*}[h!]
\centering
\small
\caption{The CNSFR characteristics.}
\label{tab:seleccion}
\begin{tabular}{cccc}
\hline
Region ID & \begin{tabular}[c]{@{}c@{}} Area \\ (arcsec$^2$)\end{tabular} & \begin{tabular}[c]{@{}c@{}}Offsets from galaxy center $^a$ \\ (arcsec)\end{tabular} & \begin{tabular}[c]{@{}c@{}}F(H$\alpha$)\\ (10$^{-15}$ erg$\cdot s^{-1}\cdot cm^{-2}$)\end{tabular} \\ \hline
R1	&	5.76	&	-5.8, 4.8	&	346.090 $\pm$ 0.525\\
R2	&	7.72	&	0.0, 9.2	&	363.548 $\pm$ 0.568\\
R3	&	7.40	&	7.0, 7.4	&	221.728 $\pm$ 0.445\\
R4	&	5.16	&	8.8, 5.4	&	203.510 $\pm$ 0.346\\
R5	&	9.92	&	-6.2, 0.8	&	300.696 $\pm$ 0.689\\
R6	&	5.12	&	10.4, -2.6	&	140.083 $\pm$ 0.268\\
R7	&	7.36	&	-4.6, -4.2	&	181.884 $\pm$ 0.394\\
R8	&	1.44	&	9.4, 3.6	&	59.412 $\pm$ 0.092\\
R9	&	4.52	&	-3.8, -6.8	&	105.152 $\pm$ 0.217\\
R10	&	5.80	&	0.8, -10.0	&	117.413 $\pm$ 0.293\\
R11	&	4.04	&	5.6, -9.8	&	106.157 $\pm$ 0.169\\
R12	&	2.76	&	3.6, -10.2	&	81.547 $\pm$ 0.139\\
R13	&	5.16	&	5.2, 9.4	&	131.487 $\pm$ 0.275\\
R14	&	4.36	&	8.0, -9.0	&	94.847 $\pm$ 0.168\\
R15	&	0.96	&	-7.4, -1.2	&	25.468 $\pm$ 0.066\\
R16	&	3.56	&	3.6, 7.8	&	64.546 $\pm$ 0.194\\
R17	&	2.28	&	-8.2, 2.4	&	42.854 $\pm$ 0.095\\
R18	&	0.80	&	0.0, -8.4	&	17.077 $\pm$ 0.041\\
R19	&	0.92	&	10.2, -7.4	&	13.287 $\pm$ 0.025\\
R20	&	1.00	&	-6.6, -4.4	&	15.571 $\pm$ 0.039\\
R21	&	2.72	&	-2.8, 11.4	&	36.827 $\pm$ 0.093\\
R22	&	0.36	&	-2.2, -7.2	&	5.519 $\pm$ 0.015\\
R23	&	0.80	&	11.0, 2.0	&	10.586 $\pm$ 0.032\\
R24	&	0.36	&	12.2, 2.8	&	3.125 $\pm$ 0.011\\ \hline
\end{tabular}
\begin{tablenotes}
\centering
\item $^a$ Offsets from centre of the galaxy to the centre of each individual region.
\end{tablenotes}
\end{table*}

\begin{table*}[h!]
\centering
\caption{Reddening corrected emission line intensities.}
\label{tab:lines}
\small
\begin{tabular}{ccccccccccc}
\hline
& Line & Hb & [OIII] & [OIII] &  [NII] & H$\alpha$ & [NII] & [SII] & [SII] & [SIII]\\
& $\lambda$ & 4861 & 4959 & 5007  & 6548 & 6563 & 6584 & 6717 & 6731 & 9069 \\
& f($\lambda$) & 0.000 & -0.024 & -0.035 & -0.311 & -0.313 & -0.316 & -0.334 & -0.336 & -0.561  \\ \hline
\multicolumn{1}{c}{Region ID} & \multicolumn{1}{c}{c(H$\beta$)} & \multicolumn{1}{c}{I(H$\beta$)$^a$}& &&& I($\lambda $)$^b$\\ \hline
R1 & 0.59 $\pm$ 0.01 & 306.96 $\pm$ 8.88 & 34 $\pm$ 5 & 91 $\pm$ 5 & 387 $\pm$ 2 & 2870 $\pm$ 35 & 1209 $\pm$ 4 &183 $\pm$ 1 & 168 $\pm$ 1 & 118 $\pm$ 2\\ 
R2 & 0.79 $\pm$ 0.01 & 444.95 $\pm$ 11.13 & 28 $\pm$ 4 & 72 $\pm$ 4 & 344 $\pm$ 2 & 2870 $\pm$ 30 & 1074 $\pm$ 3 & 218 $\pm$ 2 & 192 $\pm$ 2 & 78 $\pm$ 2\\ 
R3 & 0.78 $\pm$ 0.02 & 266.68 $\pm$ 10.67 & 37 $\pm$ 9 & 96 $\pm$ 8 & 346 $\pm$ 3 & 2870 $\pm$ 49 & 1094 $\pm$ 5 & 228 $\pm$ 2 & 196 $\pm$ 2 & 78 $\pm$ 3\\ 
R4 & 0.98 $\pm$ 0.01 & 333.22 $\pm$ 10.12 & 37 $\pm$ 6 & 94 $\pm$ 6 & 364 $\pm$ 2 & 2870 $\pm$ 37 & 1138 $\pm$ 4 & 210 $\pm$ 2 & 184 $\pm$ 2 & 79 $\pm$ 2\\ 
R5 & 0.77 $\pm$ 0.02 & 352.25 $\pm$ 18.22 & 36 $\pm$ 11 & 92 $\pm$ 10 & 319 $\pm$ 3 & 2870 $\pm$ 63 & 1023 $\pm$ 5 & 238 $\pm$ 3 & 205 $\pm$ 3 & 79 $\pm$ 4\\ 
R6 & 1.25 $\pm$ 0.02 & 350.80 $\pm$ 15.90 & 53 $\pm$ 11 & 130 $\pm$ 10 & 436 $\pm$ 3 & 2870 $\pm$ 55 & 1388 $\pm$ 5 & 295 $\pm$ 2 & 270 $\pm$ 2 & 81 $\pm$ 3\\ 
R7 & 1.32 $\pm$ 0.01 & 507.78 $\pm$ 16.49 & 40 $\pm$ 7 & 96 $\pm$ 6 & 340 $\pm$ 3 & 2870 $\pm$ 40 & 1067 $\pm$ 5 & 209 $\pm$ 3 & 186 $\pm$ 3 & 80 $\pm$ 2\\ 
R8 & 1.14 $\pm$ 0.01 & 125.35 $\pm$ 4.18 & 38 $\pm$ 7 & 94 $\pm$ 6 & 353 $\pm$ 2 & 2870 $\pm$ 40 & 1102 $\pm$ 3 & 231 $\pm$ 2 & 198 $\pm$ 2 & 70 $\pm$ 2\\ 
R9 & 1.05 $\pm$ 0.02 & 193.94 $\pm$ 8.09 & 35 $\pm$ 10 & 88 $\pm$ 9 & 305 $\pm$ 3 & 2870 $\pm$ 51 & 989 $\pm$ 5 & 216 $\pm$ 3 & 183 $\pm$ 2 & 71 $\pm$ 3\\ 
R10 & 0.88 $\pm$ 0.02 & 163.56 $\pm$ 9.78 & 33 $\pm$ 14 & 86 $\pm$ 13 & 298 $\pm$ 4 & 2870 $\pm$ 72 & 983 $\pm$ 5 & 241 $\pm$ 3 & 199 $\pm$ 3 & 70 $\pm$ 4\\ 
R11 & 1.14 $\pm$ 0.01 & 225.70 $\pm$ 8.04 & 32 $\pm$ 8 & 79 $\pm$ 8 & 319 $\pm$ 2 & 2870 $\pm$ 43 & 1015 $\pm$ 3 & 238 $\pm$ 2 & 200 $\pm$ 2 & 65 $\pm$ 2\\ 
R12 & 1.04 $\pm$ 0.02 & 147.51 $\pm$ 5.88 & 36 $\pm$ 9 & 90 $\pm$ 8 & 303 $\pm$ 2 & 2870 $\pm$ 48 & 967 $\pm$ 4 & 233 $\pm$ 2 & 199 $\pm$ 2 & 79 $\pm$ 3\\ 
R13 & 0.90 $\pm$ 0.02 & 190.78 $\pm$ 8.56 & 40 $\pm$ 10 & 101 $\pm$ 9 & 321 $\pm$ 3 & 2870 $\pm$ 54 & 1021 $\pm$ 5 & 251 $\pm$ 2 & 208 $\pm$ 2 & 49 $\pm$ 3\\ 
R14 & 0.98 $\pm$ 0.02 & 155.05 $\pm$ 5.85 & 48 $\pm$ 9 & 120 $\pm$ 8 & 305 $\pm$ 2 & 2870 $\pm$ 46 & 968 $\pm$ 4 & 266 $\pm$ 2 & 220 $\pm$ 2 & 62 $\pm$ 2\\ 
R15 & 0.67 $\pm$ 0.02 & 25.43 $\pm$ 1.45 & 37 $\pm$ 12 & 98 $\pm$ 12 & 273 $\pm$ 4 & 2870 $\pm$ 69 & 892 $\pm$ 5 & 210 $\pm$ 3 & 169 $\pm$ 3 & 77 $\pm$ 4\\ 
R16 & 0.61 $\pm$ 0.03 & 58.62 $\pm$ 3.53 & 34 $\pm$ 14 & 91 $\pm$ 13 & 292 $\pm$ 5 & 2870 $\pm$ 73 & 954 $\pm$ 7 & 224 $\pm$ 4 & 180 $\pm$ 4 & 42 $\pm$ 5\\ 
R17 & 0.90 $\pm$ 0.02 & 62.45 $\pm$ 3.26 & 38 $\pm$ 12 & 98 $\pm$ 10 & 319 $\pm$ 3 & 2870 $\pm$ 63 & 1027 $\pm$ 5 & 228 $\pm$ 2 & 183 $\pm$ 2 & 75 $\pm$ 3\\ 
R18 & 0.95 $\pm$ 0.02 & 26.90 $\pm$ 1.32 & 35 $\pm$ 11 & 89 $\pm$ 10 & 277 $\pm$ 3 & 2870 $\pm$ 60 & 901 $\pm$ 5 & 210 $\pm$ 3 & 177 $\pm$ 3 & 61 $\pm$ 4\\ 
R19 & 1.21 $\pm$ 0.02 & 31.57 $\pm$ 1.37 & 48 $\pm$ 9 & 118 $\pm$ 8 & 412 $\pm$ 3 & 2870 $\pm$ 53 & 1302 $\pm$ 5 & 230 $\pm$ 2 & 192 $\pm$ 2 & 104 $\pm$ 2\\ 
R20 & 1.28 $\pm$ 0.02 & 40.88 $\pm$ 2.40 & 67 $\pm$ 14 & 162 $\pm$ 12 & 320 $\pm$ 4 & 2870 $\pm$ 71 & 1049 $\pm$ 6 & 244 $\pm$ 3 & 200 $\pm$ 3 & 89 $\pm$ 4\\ 
R21 & 0.61 $\pm$ 0.02 & 33.85 $\pm$ 1.41 & 49 $\pm$ 9 & 131 $\pm$ 8 & 262 $\pm$ 4 & 2870 $\pm$ 51 & 833 $\pm$ 5 & 228 $\pm$ 3 & 174 $\pm$ 3 & 52 $\pm$ 4\\ 
R22 & 1.15 $\pm$ 0.02 & 11.91 $\pm$ 0.69 & 75 $\pm$ 14 & 185 $\pm$ 12 & 450 $\pm$ 4 & 2870 $\pm$ 70 & 1450 $\pm$ 7 & 300 $\pm$ 4 & 262 $\pm$ 4 & 80 $\pm$ 4\\ 
R23 & 1.35 $\pm$ 0.03 & 31.02 $\pm$ 2.28 & 78 $\pm$ 19 & 188 $\pm$ 16 & 366 $\pm$ 5 & 2870 $\pm$ 89 & 1187 $\pm$ 7 & 282 $\pm$ 4 & 232 $\pm$ 4 & 39 $\pm$ 4\\ 
R24 & 1.31 $\pm$ 0.03 & 8.62 $\pm$ 0.71 & 57 $\pm$ 20 & 138 $\pm$ 17 & 352 $\pm$ 5 & 2870 $\pm$ 99 & 1159 $\pm$ 8 & 248 $\pm$ 4 & 201 $\pm$ 4 & 60 $\pm$ 4\\ \hline
\end{tabular}
\begin{tablenotes}
\centering
\item $^a$ In units of 10$^{-15}$ erg/s/cm$^2$.\
\item $^b$ Values normalized to I(H$\beta$) 10$^{-3}$. 
\end{tablenotes}
\end{table*}

\begin{table*}[h!]
\centering
\caption{Ionic and total sulphur abundances derived by the direct method for the CNSFRs with measured [SIII]$\lambda$ 6312 \AA\ line intensities.}
\label{tab:sulfur_measurements}
\begin{tabular}{ccccccc}
\hline
Region ID & I([SIII]$\lambda $6312)$^a$& R$_{S3}$& t$_e$([SIII])$^b$ & 12+log(S$^{+}$/H$^{+}$)&	12+log(S$^{++}$/H$^{+}$) & 12+log(S/H)\\ \hline
R1	 & 	196.7 $\pm$ 6.5	 & 	631.4 $\pm$ 107.0	 & 	0.4807 $\pm$ 0.0254	 & 	7.0532 $\pm$ 0.0979	 & 	7.1554 $\pm$ 0.2682	 & 	7.408 $\pm$ 0.156\\ 
R2	 & 	448.2 $\pm$ 12.9	 & 	266.9 $\pm$ 48.1	 & 	0.5957 $\pm$ 0.0240	 & 	6.7205 $\pm$ 0.0604	 & 	6.6680 $\pm$ 0.1653	 & 	6.996 $\pm$ 0.084\\ 
R3	 & 	107.4 $\pm$ 5.7	 & 	668.9 $\pm$ 168.9	 & 	0.4719 $\pm$ 0.0391	 & 	7.1736 $\pm$ 0.1562	 & 	7.0079 $\pm$ 0.4281	 & 	7.400 $\pm$ 0.197\\ 
R4	 & 	88.8 $\pm$ 4.2	 & 	1018.8 $\pm$ 348.0	 & 	0.3970 $\pm$ 0.0705	 & 	7.5304 $\pm$ 0.3972	 & 	7.3071 $\pm$ 1.0898	 & 	7.734 $\pm$ 0.478\\ 
R6	 & 	93.0 $\pm$ 5.4	 & 	1047.7 $\pm$ 281.5	 & 	0.3912 $\pm$ 0.0567	 & 	7.7239 $\pm$ 0.3289	 & 	7.3448 $\pm$ 0.9026	 & 	7.875 $\pm$ 0.353\\ 
R8	 & 	77.0 $\pm$ 3.0	 & 	392.0 $\pm$ 76.0	 & 	0.5459 $\pm$ 0.0249	 & 	6.8933 $\pm$ 0.0746	 & 	6.7398 $\pm$ 0.2045	 & 	7.124 $\pm$ 0.095\\ 
R11	 & 	131.5 $\pm$ 6.4	 & 	383.8 $\pm$ 112.6	 & 	0.5486 $\pm$ 0.0377	 & 	6.8943 $\pm$ 0.1116	 & 	6.7006 $\pm$ 0.3056	 & 	7.109 $\pm$ 0.137\\ 
R12	 & 	126.1 $\pm$ 6.3	 & 	316.8 $\pm$ 76.6	 & 	0.5733 $\pm$ 0.0311	 & 	6.8098 $\pm$ 0.0845	 & 	6.7227 $\pm$ 0.2315	 & 	7.069 $\pm$ 0.114\\ 
R14	 & 	135.8 $\pm$ 6.7	 & 	244.4 $\pm$ 65.5	 & 	0.6075 $\pm$ 0.0367	 & 	6.7617 $\pm$ 0.0888	 & 	6.5430 $\pm$ 0.2430	 & 	6.967 $\pm$ 0.107\\ 
R19	 & 	39.1 $\pm$ 2.0	 & 	288.8 $\pm$ 68.2	 & 	0.5853 $\pm$ 0.0309	 & 	6.7632 $\pm$ 0.0805	 & 	6.8155 $\pm$ 0.2200	 & 	7.091 $\pm$ 0.123\\ 
R21	 & 	75.8 $\pm$ 3.8	 & 	79.4 $\pm$ 18.4	 & 	0.8302 $\pm$ 0.0697	 & 	6.2239 $\pm$ 0.0912	 & 	6.0980 $\pm$ 0.2491	 & 	6.467 $\pm$ 0.119\\ 
\hline
\end{tabular}
\begin{tablenotes}
\centering
\item $^a$ In units of 10$^{-18}$ erg/s/cm$^2$.\\
\item $^b$ In units of 10$^{4}$ K.\\
\end{tablenotes}
\end{table*}

\begin{table}[h!]
\centering
\caption{Sulphur abundances of the observed CNSFRs derived by empirical methods.}
\label{tab3}
\begin{tabular}{ccc}
\hline
Region ID & S23 &  12+log(S/H)\\ \hline
R1 & 0.756 $\pm$ 0.008 & 6.384 $\pm$ 0.016\\ 
R2 & 0.679 $\pm$ 0.008 & 6.295 $\pm$ 0.017\\ 
R3 & 0.694 $\pm$ 0.011 & 6.313 $\pm$ 0.019\\ 
R4 & 0.665 $\pm$ 0.008 & 6.279 $\pm$ 0.017\\ 
R5 & 0.714 $\pm$ 0.014 & 6.337 $\pm$ 0.021\\ 
R6 & 0.843 $\pm$ 0.01 & 6.479 $\pm$ 0.016\\ 
R7 & 0.669 $\pm$ 0.009 & 6.284 $\pm$ 0.018\\ 
R8 & 0.669 $\pm$ 0.007 & 6.284 $\pm$ 0.017\\ 
R9 & 0.643 $\pm$ 0.012 & 6.252 $\pm$ 0.021\\ 
R10 & 0.681 $\pm$ 0.014 & 6.298 $\pm$ 0.022\\ 
R11 & 0.662 $\pm$ 0.008 & 6.276 $\pm$ 0.018\\ 
R12 & 0.703 $\pm$ 0.009 & 6.324 $\pm$ 0.017\\ 
R13 & 0.629 $\pm$ 0.011 & 6.235 $\pm$ 0.021\\ 
R14 & 0.7 $\pm$ 0.009 & 6.32 $\pm$ 0.017\\ 
R15 & 0.643 $\pm$ 0.015 & 6.253 $\pm$ 0.024\\ 
R16 & 0.549 $\pm$ 0.018 & 6.134 $\pm$ 0.029\\ 
R17 & 0.668 $\pm$ 0.012 & 6.283 $\pm$ 0.02\\ 
R18 & 0.597 $\pm$ 0.013 & 6.196 $\pm$ 0.023\\ 
R19 & 0.779 $\pm$ 0.009 & 6.41 $\pm$ 0.016\\ 
R20 & 0.748 $\pm$ 0.014 & 6.375 $\pm$ 0.021\\ 
R21 & 0.58 $\pm$ 0.013 & 6.174 $\pm$ 0.024\\ 
R22 & 0.836 $\pm$ 0.015 & 6.471 $\pm$ 0.019\\ 
R23 & 0.646 $\pm$ 0.015 & 6.257 $\pm$ 0.024\\ 
R24 & 0.657 $\pm$ 0.016 & 6.269 $\pm$ 0.024\\ 
\hline
\end{tabular}
\end{table}

\begin{table*}
\centering
\caption{Characteristics of the observed CNSFRs.}
\label{tab:HIIcharacteristics}
\begin{tabular}{ccccccc}
\hline
Region ID &\begin{tabular}[c]{@{}c@{}}L(H$\alpha$) \\ (erg s$^{-1}$)\end{tabular}& \begin{tabular}[c]{@{}c@{}}Q(H$_0$) \\ (photons s$^{-1}$)\end{tabular}&\begin{tabular}[c]{@{}c@{}}log(u) \\ \end{tabular}&\begin{tabular}[c]{@{}c@{}}n$_e$ \\ (cm$^{-3}$)\end{tabular}& \begin{tabular}[c]{@{}c@{}}M(HII) \\ (M$_\odot$)\end{tabular}& \begin{tabular}[c]{@{}c@{}}M$_{ion}$ \\ (M$_\odot$)\end{tabular}\\ \hline
R1 & (221.7 $\pm$ 6.4) $\times$ 10$^{38}$ & (162.2 $\pm$ 4.7) $\times$ 10$^{50}$ & -2.887 $\pm$ 0.034 & 385 $\pm$ 37 & (13.4 $\pm$ 1.5) $\times$ 10$^{4}$ & (44.2 $\pm$ 3.4) $\times$ 10$^{5}$\\ 
R2 & (32.1 $\pm$ 2.5) $\times$ 10$^{39}$ & (235.1 $\pm$ 5.9) $\times$ 10$^{50}$ & -2.763 $\pm$ 0.030 & 313 $\pm$ 32 & (24.0 $\pm$ 2.2) $\times$ 10$^{4}$ & (49.4 $\pm$ 3.8) $\times$ 10$^{5}$\\ 
R3 & (19.3 $\pm$ 1.6) $\times$ 10$^{39}$ & (140.9 $\pm$ 5.6) $\times$ 10$^{50}$ & -2.907 $\pm$ 0.033 & 272 $\pm$ 34 & (16.5 $\pm$ 1.7) $\times$ 10$^{4}$ & (52.9 $\pm$ 4.2) $\times$ 10$^{5}$\\ 
R4 & (24.1 $\pm$ 1.9) $\times$ 10$^{39}$ & (176.1 $\pm$ 5.4) $\times$ 10$^{50}$ & -2.695 $\pm$ 0.036 & 299 $\pm$ 33 & (18.8 $\pm$ 2.1) $\times$ 10$^{4}$ & (47.7 $\pm$ 3.7) $\times$ 10$^{5}$\\ 
R5 & (25.4 $\pm$ 2.3) $\times$ 10$^{39}$ & (186.1 $\pm$ 9.6) $\times$ 10$^{50}$ & -2.915 $\pm$ 0.033 & 274 $\pm$ 38 & (21.7 $\pm$ 2.1) $\times$ 10$^{4}$ & (80.8 $\pm$ 7.1) $\times$ 10$^{5}$\\ 
R6 & (25.3 $\pm$ 2.2) $\times$ 10$^{39}$ & (185.4 $\pm$ 8.4) $\times$ 10$^{50}$ & -2.768 $\pm$ 0.039 & 376 $\pm$ 37 & (15.7 $\pm$ 1.9) $\times$ 10$^{4}$ & (83.8 $\pm$ 6.9) $\times$ 10$^{5}$\\ 
R7 & (36.7 $\pm$ 2.9) $\times$ 10$^{39}$ & (268.3 $\pm$ 8.8) $\times$ 10$^{50}$ & -2.698 $\pm$ 0.032 & 322 $\pm$ 45 & (26.6 $\pm$ 2.6) $\times$ 10$^{4}$ & (92.7 $\pm$ 7.0) $\times$ 10$^{5}$\\ 
R8 & (90.5 $\pm$ 7.2) $\times$ 10$^{38}$ & (66.2 $\pm$ 2.2) $\times$ 10$^{50}$ & -2.505 $\pm$ 0.066 & 261 $\pm$ 26 & (8.1 $\pm$ 1.7) $\times$ 10$^{4}$ & (20.5 $\pm$ 1.6) $\times$ 10$^{5}$\\ 
R9 & (14.0 $\pm$ 1.2) $\times$ 10$^{39}$ & (102.5 $\pm$ 4.3) $\times$ 10$^{50}$ & -2.787 $\pm$ 0.041 & 246 $\pm$ 36 & (13.3 $\pm$ 1.7) $\times$ 10$^{4}$ & (42.2 $\pm$ 3.4) $\times$ 10$^{5}$\\ 
R10 & (11.8 $\pm$ 1.1) $\times$ 10$^{39}$ & (86.4 $\pm$ 5.2) $\times$ 10$^{50}$ & -2.897 $\pm$ 0.041 & 208 $\pm$ 35 & (13.2 $\pm$ 1.6) $\times$ 10$^{4}$ & (44.4 $\pm$ 4.1) $\times$ 10$^{5}$\\ 
R11 & (16.3 $\pm$ 1.3) $\times$ 10$^{39}$ & (119.3 $\pm$ 4.3) $\times$ 10$^{50}$ & -2.650 $\pm$ 0.041 & 234 $\pm$ 25 & (16.3 $\pm$ 2.1) $\times$ 10$^{4}$ & (45.5 $\pm$ 3.5) $\times$ 10$^{5}$\\ 
R12 & (106.5 $\pm$ 8.8) $\times$ 10$^{38}$ & (78.0 $\pm$ 3.1) $\times$ 10$^{50}$ & -2.712 $\pm$ 0.049 & 258 $\pm$ 28 & (9.6 $\pm$ 1.5) $\times$ 10$^{4}$ & (26.0 $\pm$ 2.1) $\times$ 10$^{5}$\\ 
R13 & (13.8 $\pm$ 1.2) $\times$ 10$^{39}$ & (100.8 $\pm$ 4.5) $\times$ 10$^{50}$ & -2.785 $\pm$ 0.039 & 211 $\pm$ 28 & (15.2 $\pm$ 1.8) $\times$ 10$^{4}$ & (40.3 $\pm$ 3.3) $\times$ 10$^{5}$\\ 
R14 & (112.0 $\pm$ 9.2) $\times$ 10$^{38}$ & (81.9 $\pm$ 3.1) $\times$ 10$^{50}$ & -2.803 $\pm$ 0.040 & 211 $\pm$ 24 & (12.4 $\pm$ 1.6) $\times$ 10$^{4}$ & (29.8 $\pm$ 2.4) $\times$ 10$^{5}$\\ 
R15 & (18.4 $\pm$ 1.7) $\times$ 10$^{38}$ & (134.4 $\pm$ 7.7) $\times$ 10$^{49}$ & -2.851 $\pm$ 0.082 & 176 $\pm$ 36 & (24.3 $\pm$ 6.4) $\times$ 10$^{3}$ & (54.2 $\pm$ 5.1) $\times$ 10$^{4}$\\ 
R16 & (42.3 $\pm$ 4.0) $\times$ 10$^{38}$ & (31.0 $\pm$ 1.9) $\times$ 10$^{50}$ & -3.045 $\pm$ 0.049 & 171 $\pm$ 43 & (57.8 $\pm$ 8.4) $\times$ 10$^{3}$ & (14.3 $\pm$ 1.3) $\times$ 10$^{5}$\\ 
R17 & (45.1 $\pm$ 4.0) $\times$ 10$^{38}$ & (33.0 $\pm$ 1.7) $\times$ 10$^{50}$ & -2.822 $\pm$ 0.056 & 170 $\pm$ 28 & (6.2 $\pm$ 1.1) $\times$ 10$^{4}$ & (12.5 $\pm$ 1.1) $\times$ 10$^{5}$\\ 
R18 & (19.4 $\pm$ 1.7) $\times$ 10$^{38}$ & (142.2 $\pm$ 7.0) $\times$ 10$^{49}$ & -2.877 $\pm$ 0.089 & 237 $\pm$ 44 & (19.1 $\pm$ 5.4) $\times$ 10$^{3}$ & (58.2 $\pm$ 5.0) $\times$ 10$^{4}$\\ 
R19 & (22.8 $\pm$ 1.9) $\times$ 10$^{38}$ & (166.8 $\pm$ 7.3) $\times$ 10$^{49}$ & -2.838 $\pm$ 0.082 & 221 $\pm$ 27 & (24.0 $\pm$ 6.4) $\times$ 10$^{3}$ & (53.2 $\pm$ 4.5) $\times$ 10$^{4}$\\ 
R20 & (29.5 $\pm$ 2.8) $\times$ 10$^{38}$ & (21.6 $\pm$ 1.3) $\times$ 10$^{50}$ & -2.709 $\pm$ 0.081 & 196 $\pm$ 36 & (35.2 $\pm$ 9.1) $\times$ 10$^{3}$ & (93.0 $\pm$ 8.7) $\times$ 10$^{4}$\\ 
R21 & (24.4 $\pm$ 2.0) $\times$ 10$^{38}$ & (178.9 $\pm$ 7.5) $\times$ 10$^{49}$ & -2.961 $\pm$ 0.050 & 107 $\pm$ 29 & (53.5 $\pm$ 8.4) $\times$ 10$^{3}$ & (58.9 $\pm$ 4.9) $\times$ 10$^{4}$\\ 
R22 & (86.0 $\pm$ 8.0) $\times$ 10$^{37}$ & (63.0 $\pm$ 3.6) $\times$ 10$^{49}$ & -2.979 $\pm$ 0.131 & 295 $\pm$ 46 & (6.8 $\pm$ 2.9) $\times$ 10$^{3}$ & (28.8 $\pm$ 2.7) $\times$ 10$^{4}$\\ 
R23 & (22.4 $\pm$ 2.3) $\times$ 10$^{38}$ & (16.4 $\pm$ 1.2) $\times$ 10$^{50}$ & -2.739 $\pm$ 0.092 & 199 $\pm$ 38 & (26.3 $\pm$ 7.6) $\times$ 10$^{3}$ & (9.9 $\pm$ 1.0) $\times$ 10$^{5}$\\ 
R24 & (62.2 $\pm$ 6.8) $\times$ 10$^{37}$ & (45.5 $\pm$ 3.7) $\times$ 10$^{49}$ & -2.914 $\pm$ 0.133 & 184 $\pm$ 42 & (7.9 $\pm$ 3.4) $\times$ 10$^{3}$ & (24.0 $\pm$ 2.7) $\times$ 10$^{4}$\\ \hline
\end{tabular}
\end{table*}

\begin{table}
\small
\caption{Comparison between our results and those from \citet[][K00]{Kotilainen2000} and \citet[][H87]{Hummel1987}, in infrared and radio wavelengths respectively.}\label{tab:multiw}
\begin{tabular}{ccccccccccccccccc}
\cline{4-17} 
 &      & & \multicolumn{3}{c}{Excitation} & \multicolumn{2}{c}{A$_V$} & \multicolumn{3}{c}{Q(H) [$10^{52}$ phot/s]} & \multicolumn{2}{c}{SFR [M$_\odot$/yr]} & \multicolumn{2}{c}{Mass [$10^{6}$]} & \multicolumn{2}{c}{Age [Ma]} \\
ID & K00                  & H87  & [OIII]/[NII]  & H$_2$/Bra  & S$_{th}$/S$_{tot}$ & Opt     & IR       & Opt        & IR            & Radio         & Opt      & IR       & Opt     & IR         & Opt       & IR       \\ \hline
R1        & K1                     &           & -0.66     & 0.61    &          & 1.28        & 1.00     & 1.62       & 2.82      &               & 0.18         & 0.95     & 4.42    & 6.56   & 5.68    & 6.23     \\
R2        & K3,4                   & H1        & -0.61     &         & 0.56     & 1.73        & 1.30     & 2.35       & 2.26      & 8.30      & 0.25         & 1.04     & 4.94    & 6.59   & 5.56    & 6.36    \\
R3        & K5                     &           & -0.62     & 0.58    &          & 1.70        & 1.20     & 1.41       & 0.58      &               & 0.15         & 0.32     & 5.29    & 2.02   & 5.85    & 6.53     \\
R4        & K6                     & H10       & -0.64     & 0.32    & 0.48     & 2.13        & 1.50     & 1.76       & 0.78      & 5.70      & 0.19         & 0.37     & 4.77    & 2.33   & 5.68    & 6.37     \\
R5        & K14                    & H5        & -0.59     & 0.77    & 0.44     & 1.67        & 2.20     & 1.86       & 2.50      & 6.50      & 0.20         & 1.32     & 8.08    & 8.51   & 5.96    & 6.46     \\
R6        & \multicolumn{1}{r}{K8} & H8        & -0.72     & 0.91    & 0.45     & 2.71        & 1.80     & 1.85       & 2.01      & 7.10      & 0.20         & 1.00     & 8.38    & 6.39   & 5.99     & 6.4      \\
R7        & K13                    & H5        & -0.61     & 0.36    & 0.44     & 2.86        & 2.20     & 2.68       & 2.00      & 6.50      & 0.29         & 0.93     & 9.27    & 5.89   & 5.78    & 6.36     \\
R8        & K7                     & H9        & -0.62     & 0.63    & 0.02     & 2.47        & 1.50     & 0.66       & 0.52      &               & 0.07         & 0.28     & 2.05    & 1.82   & 5.73    & 6.51     \\
R9        &                        & H6        & -0.58     &         & 0.3      & 2.29        &          & 1.02       &               & 4.60      & 0.11         &          & 4.22    &            & 5.92    &          \\
R12       & K11                    &           & -0.57     & 0.52    &          & 2.26        & 0.10     & 0.78       & 0.49      &               & 0.08         & 0.27     & 2.60    & 1.79   & 5.75    & 6.57     \\
R14       & K10                    &           & -0.57     & 1.36    &          & 2.49        &          & 0.82       & 0.55     &               & 0.09         & 0.34     & 2.98    & 2.29   & 5.83    & 6.65    \\ \hline

\end{tabular}
\end{table}

\end{document}